\documentclass[preprint,aps,jcp,onecolumn,superscriptaddress]{revtex4-1}
\usepackage{graphicx}
\usepackage{mhchem}
\usepackage[T1]{fontenc} % Use modern font encodings
\usepackage{enumerate}

\begin{document}
	
	\title{O1NumHess: a fast and accurate seminumerical Hessian algorithm using only O(1) gradients}
	
	\author{Bo Wang}
	\affiliation{Qingdao Institute for Theoretical and Computational Sciences, Center for Optics Research and Engineering, Shandong University, Qingdao, Shandong 266237, P. R. China}
	
	\author{Shaohang Luo}
	\affiliation{College of Mathematics and Statistics, Chongqing University, Chongqing 401331, P. R. China}
	
	\author{Zikuan Wang}\email{wzkchem5@sdu.edu.cn}
	\affiliation{Qingdao Institute for Theoretical and Computational Sciences, Center for Optics Research and Engineering, Shandong University, Qingdao, Shandong 266237, P. R. China}
	
	\author{Wenjian Liu}\email{liuwj@sdu.edu.cn}
	\affiliation{Qingdao Institute for Theoretical and Computational Sciences, Center for Optics Research and Engineering, Shandong University, Qingdao, Shandong 266237, P. R. China}
	
	\begin{abstract}
		In this work, we describe a new algorithm, O1NumHess, to calculate the Hessian of a molecular system by finite differentiation of gradients calculated at displaced geometries. Different from the conventional seminumerical Hessian algorithm, which requires gradients at $O(N_{\mathrm{atom}})$ displaced geometries (where $N_{\mathrm{atom}}$ is the number of atoms), the present approach only requires $O(1)$ gradients. Key to the reduction of the number of gradients is the exploitation of the off-diagonal low-rank (ODLR) property of Hessians, namely the blocks of the Hessian that correspond to two distant groups of atoms have low rank. This property reduces the number of independent entries of the Hessian from $O(N_{\mathrm{atom}}^2)$ to $O(N_{\mathrm{atom}})$, such that $O(1)$ gradients already contain enough information to uniquely determine the Hessian. Numerical results on model systems (long alkanes and polyenes), transition metal reactions (WCCR10) and non-covalent complexes (S30L-CI) using the BDF program show that O1NumHess gives frequency, zero-point energy, enthalpy and Gibbs free energy errors that are only about two times those of conventional double-sided seminumerical Hessians. Moreover, O1NumHess is always faster than the conventional numerical Hessian algorithm, frequently even faster than the analytic Hessian, and requires only about 100 gradients for sufficiently large systems. An open-source implementation of this method, which can also be applied to problems irrelevant to computational chemistry, is available on GitHub.
	\end{abstract}
	
	\maketitle
	
	\section{Introduction}\label{Introduction}
	
	The Hessian of a twice differentiable multivariate function $E(\xi_1,\ldots,\xi_n)$, i.e.~the second order derivative matrix
	$H_{ij} \equiv \frac{\partial ^2 E(\xi_1,\ldots,\xi_n)}{\partial \xi_i \partial \xi_j}$,
	plays an important role in characterizing the behavior of the function in the neighborhood of a prescribed set of variables $(\xi_1,\ldots,\xi_n)$.
	In computational chemistry, the word ``Hessian'' by default means the nuclear Hessian of the electronic energy, where $\xi_i$ is the $i$-th nuclear coordinate (without loss of generality, we denote $\xi_{3j-2}$ ($\xi_{3j-1}$, $\xi_{3j}$) as the $x$ ($y$, $z$) coordinate of the $j$-th atom, where $j=1,2,\ldots$), $E(\xi_1,\ldots,\xi_n)$ is the electronic energy of the molecule at the coordinates $(\xi_1,\ldots,\xi_n)$, and the number of coordinates $n$ is three times the number of atoms $N_{\mathrm{atom}}$. The Hessian of a molecule is useful in a variety of ways. For example, diagonalizing the mass-weighted Hessian evaluated at a stationary structure yields the harmonic vibrational frequencies (and from that, infrared (IR) and Raman spectra), and the number of imaginary frequencies determines whether the structure is a local minimum, a transition state, or a high-order saddle point. Thermochemical quantities, such as zero-point energy (ZPE), enthalpy, entropy and Gibbs free energy, can be calculated from the vibrational frequencies under the harmonic approximation\cite{Shermo}. Hessians are also routinely evaluated along reaction paths, such as relaxed scan or intrinsic reaction coordinate (IRC) trajectories, as parts of variational transition state theory (VTST) or tunneling coefficient calculations\cite{VTST,tunneling}, to name a few. Transition rates and vibrationally resolved absorption/emission spectra between two electronic states can be calculated from the Hessians of both states, through either an enumeration of vibronic levels\cite{FCclasses1,FCclasses2} or time domain propagation\cite{MOMAP, TVCF, ESD}. Finally, Hessians calculated at general non-stationary structures can be used to speed up geometry optimization from these structures to nearby stationary points, via the Newton-Raphson method and its generalizations.
	
	A Hessian can be calculated analytically, seminumerically (i.e.~via first order finite difference of the analytic gradient), or fully numerically (via second-order finite difference of the energy). When analytic gradients are available, there is no benefit of evaluating the Hessian fully numerically, and meanwhile, the analytic Hessian is usually both more accurate (due to the lack of finite difference errors) and cheaper to calculate than the seminumerical Hessian. The high computational cost of seminumerical Hessians is due to the need of gradient calculations at a minimum of $3N_{\mathrm{atom}}+1$ displaced geometries (or $6N_{\mathrm{atom}}$ displaced geometries, if double-sided finite difference is used instead of single-sided finite difference).
	However, analytic Hessians are relatively hard to implement\cite{HFHessian, DFTHessian, ORCAHessian}, especially for methods like TDDFT\cite{TDAHessian, TDDFTHessian} and post-Hartree-Fock methods\cite{MP2Hessian, CCSDHessian, CCSDTHessian}. Moreover, they tend to take up much more memory than numerical Hessians, due to the need of solving a coupled-perturbed self-consistent field (CP-SCF) equation with $3N_{\mathrm{atom}}$ right hand sides. For molecules with a few hundreds of atoms, it is not uncommon for an analytic Hessian calculation to fail because of insufficient memory, or take up much more time than expected because the CP-SCF equation has to be solved in multiple batches to accommodate the limited available memory. Finally, numerical Hessians can be embarrassingly parallelized, while the efficient parallelization of analytic Hessians is far from trivial\cite{NUMFREQGrid}. Therefore, seminumerical Hessians are still commonly used in computational chemistry, especially when the analytic Hessian implementation is not available, poorly parallelized, or memory-intensive.
	
	In the last decade, it has been realized that (semi)numeric Hessians can be evaluated with less than $O(N_{\mathrm{atom}})$ gradients, or less than $O(N_{\mathrm{atom}}^2)$ energies, by assuming that the Hessian is sparse. The feasibility of these approaches can be easily seen from an information-theoretic point of view. If the Hessian matrix element $H_{ij}$ is ``local'', in the sense that the matrix element is only non-negligible when the Cartesian coordinates $i$ and $j$ belong to atoms that are spatially close, then the Hessian will have only $O(N_{\mathrm{atom}})$ non-negligible matrix elements. By comparison, each gradient contains $O(N_{\mathrm{atom}})$ numbers, while an energy is a single number. Therefore, it should be possible to recover the Hessian using $O(N_{\mathrm{atom}})$ energies or $O(1)$ gradients.
	Indeed, by only calculating the Hessian matrix elements between atoms that are within 3.2 \AA{} of each other, Lu and Bian et al.\cite{LuBian2024}~achieved a four-fold speedup compared to the conventional fully numerical Hessian algorithm already for $n$-\ce{C10H22}, while introducing a root mean square (RMS) frequency error of around 5 cm$^{-1}$. Similar techniques are known for higher-order nuclear derivatives of the electronic energy as well\cite{Rai2019,Sharada2022}.
	
	By comparison, calculating the seminumerical Hessian using $O(1)$ gradients is significantly more difficult. In the conventional seminumerical Hessian algorithm, one displaces only one nuclear coordinate for each displaced geometry. Thus, each gradient calculation (or every two gradient calculations for double-sided finite difference) yields one row of the Hessian. In this case, the Hessian cannot be extracted from less than $O(N_{\mathrm{atom}})$ gradients even if it is local, since otherwise there will be $O(N_{\mathrm{atom}})$ coordinates that are not perturbed even once, and it is impossible to extract the Hessian matrix element over these coordinates from the gradient data. To obtain the Hessian using $O(1)$ gradients, many or even all of the atoms should be perturbed simultaneously in any given displaced geometry. Even then, the extraction of the Hessian matrix elements from the gradient data is highly non-trivial. Using gradients computed at randomly displaced geometries, Aspuru-Guzik et al.\cite{Aspuru2015}~demonstrated that the Hessian can be recovered by compressed sensing, which only pre-supposes the sparsity of the Hessian, but does not take into account the locations of the non-negligible elements in the Hessian matrix. Not making use of the latter increases the necessary number of gradients to $O(\log N_{\mathrm{atom}})$, and the randomness of the displacement directions makes the result subject to stochastic noise.
	Alternatively, one can use $O(1)$ gradient calculations to extract a single vibrational frequency, or a small number of frequencies of a system through a Davidson-like iteration process\cite{Reiher2003,HeadGordon2014}. However, extending these approaches to calculate all the frequencies of a system (which is necessary for e.g.~calculating the thermochemical properties) would still lead to the calculation of $O(N_{\mathrm{atom}})$ gradients. Still another possibility is to use $O(1)$ gradient calculations to estimate an $O(1)$-sized sub-block of the Hessian, which is exemplified by the partial Hessian vibrational analysis (PHVA)\cite{PHVA1,PHVA2,PHVA3}, block Hessian\cite{block_hessian}, and mobile block Hessian (MBH)\cite{MBH1,MBH2,MBH3,MBH4} approaches.
	
	In this work, we present the first algorithm, hereafter termed O1NumHess, for calculating the complete seminumerical Hessian of a molecular system using $O(1)$ gradients. Apart from a logarithmic speedup compared to the compressed sensing technique\cite{Aspuru2015}, here we do not assume that the Hessian is sparse, but only use the fact that the Hessian is composed of a local component plus a low-rank component, or equivalently speaking, that the off-diagonal block $\{H_{ij}, i\in M_1, j\in M_2\}$ of the Hessian has low rank when the molecular fragment $M_1$ is distant from molecular fragment $M_2$. The paper is organized as follows. We first show numerically and theoretically that even when they are not local, Hessians still satisfy the off-diagonal low-rank (ODLR) property. Then, we introduce the algorithm of extracting the Hessian from $O(1)$ gradients using the ODLR property of the Hessian. Subsequently, we present numerical tests of the accuracy of O1NumHess vibrational frequencies and thermochemical properties of large covalent and non-covalent complexes. The paper is concluded with a discussion of future directions.
	
	\section{The ODLR property of Hessians} \label{sec:ODLR}
	
\begin{figure}
		\centering
		\includegraphics[width=\textwidth]{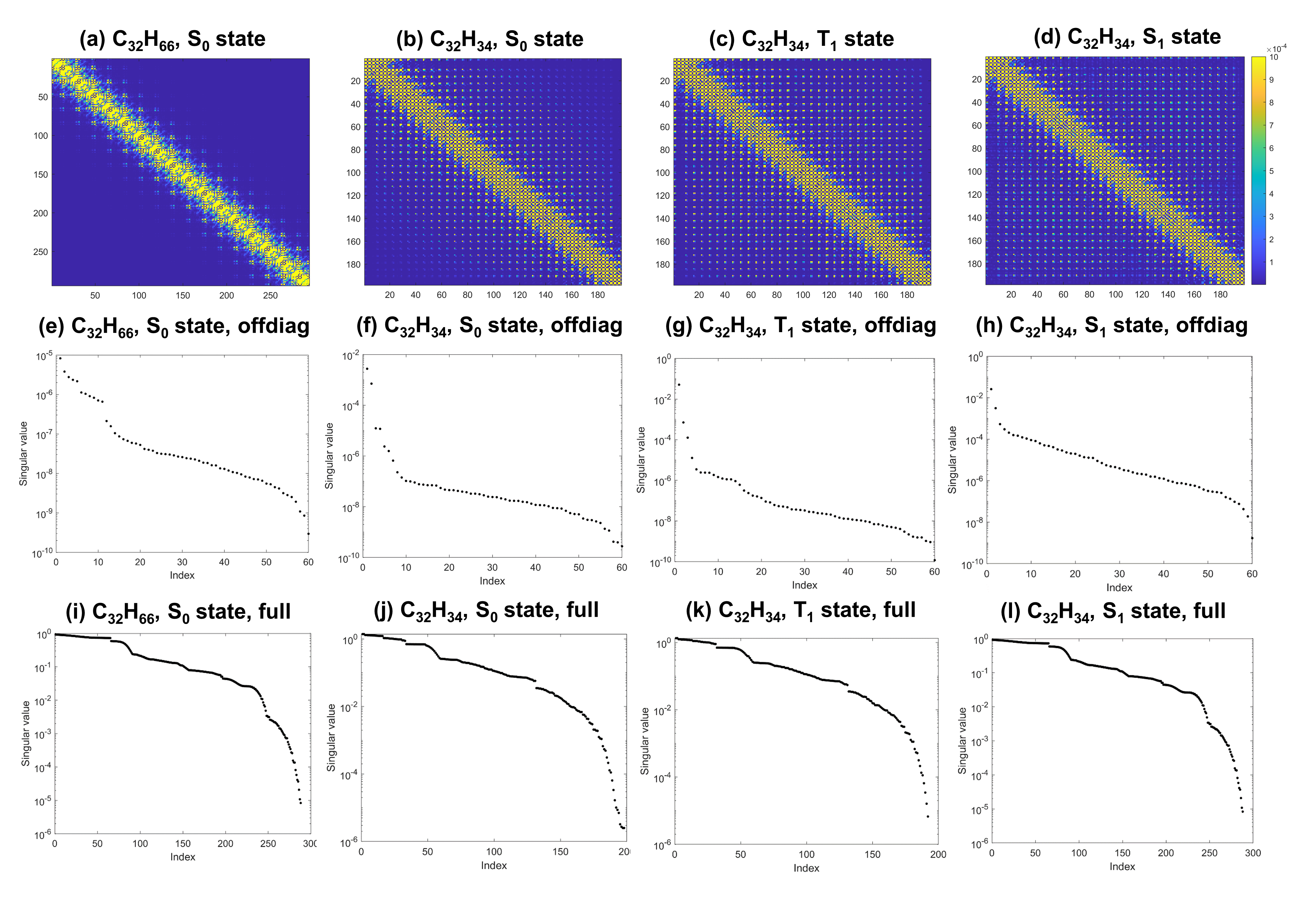}
		\caption{Heatmaps of the exact B3LYP-D3/def2-SV(P) Hessians of (a) $n$-\ce{C32H66}, as well as (b-d) the $S_0$, $S_1$ and $T_1$ states of \ce{C32H34}; (e-h) singular values of the upper right $60\times60$ blocks of the Hessians; (i-l) singular values of the whole Hessians. The matrix elements are color-coded based on their absolute values.}
		\label{heatmap}
	\end{figure}

	To begin with, let us inspect the sparsity patterns of the Hessians of two prototypical linear molecules: the linear alkane $n$-\ce{C32H66} and the linear polyene \ce{C32H34}. Both Hessians were computed at the B3LYP\cite{B88x,LYP,Becke93,B3LYP}-D3\cite{DFT-D3-1,DFT-D3-2}/def2-SV(P)\cite{def2} level of theory. As shown in Figure~\ref{heatmap}(a), the ground state Hessian of $n$-\ce{C32H66} is well approximated by a band-diagonal matrix. In contrast, while the ground state Hessian of \ce{C32H34} is dominated by a band-diagonal component (Figure~\ref{heatmap}(b)), it also possesses regular spaced non-negligible matrix elements far away from the diagonal, as illustrated by the yellow dots in the heatmap. The latter pattern is much more pronounced for the $T_1$ and $S_1$ states of \ce{C32H34} (Figure~\ref{heatmap}(c-d)), where the far off-diagonal matrix elements not only extend across the full range of the molecule, but do not appear to decay with distance. (Here and in the remaining parts of the manuscript, all triplet excited states are calculated using the unrestricted Kohn-Sham (UKS) approach, and all singlet states are calculated with full TDDFT.) These results demonstrate that while Hessians of large molecules possess some sparsity, the number of non-negligible elements is not necessarily $O(N)$, but can potentially reach $O(N^2)$, especially for excited states.
	
	Nevertheless, the regular patterns of the off-diagonal Hessian blocks suggest that there might be underlying mathematical structures within the blocks. Indeed, singular value decompositions (SVD) of the off-diagonal Hessian blocks of \ce{C32H34}, depicted in Figure~\ref{heatmap}(f-h), show that these blocks have low numerical rank, as only a small number of the singular values are non-negligible. Moreover, the rates at which the singular values decay with the singular value's index are similar for the $S_0$, $T_1$ and $S_1$ states of \ce{C32H34}, which are in turn even faster than those of $n$-\ce{C32H66} (Figure~\ref{heatmap}(e)). This contrasts with the SVD of the whole Hessians, which shows almost full numerical ranks (Figure~\ref{heatmap}(i-l); the null space consists of the translational and rotational modes and is merely 6-dimensional).
	
	The observations can be rationalized by the following argument. The electronic energy $E_I$ of the $I$-th state of a system is given by the Schr\"o{}dinger equation (the full electronic Hamiltonian $\hat{H}$ should not be confused with the Hessian $\mathbf{H}$):
	\begin{equation}
		\hat{H} |\Psi_I\rangle = E_I |\Psi_I\rangle.
	\end{equation}
	Suppose that the wavefunction $|\Psi\rangle$ does not explicitly depend on the nuclear coordinates. Then the Hellmann-Feynman theorem holds:
	\begin{equation}
		\frac{\partial E_I}{\partial \xi_j} = \langle\Psi_I| \frac{\partial \hat{H}}{\partial \xi_j} |\Psi_I\rangle. \label{HFey}
	\end{equation}
	Differentiation of Eq.~\eqref{HFey} with respect to another nuclear coordinate $\xi_i$ yields
	\begin{eqnarray}
		H_{ij} = \frac{\partial^2 E_I}{\partial \xi_i \partial \xi_j} & = & \langle\Psi_I| \frac{\partial^2 \hat{H}}{\partial \xi_i \partial \xi_j} |\Psi_I\rangle + \left(\langle\Psi_I| \frac{\partial \hat{H}}{\partial \xi_j} |\frac{\partial \Psi_I}{\partial \xi_i}\rangle + \mathrm{c.c.}\right) \label{2ndderiv0} \\
		& = & \langle\Psi_I| \frac{\partial^2 \hat{H}}{\partial \xi_i \partial \xi_j} |\Psi_I\rangle + \left(\langle\Psi_I| \frac{\partial \hat{H}}{\partial \xi_j} \hat{R} \frac{\partial \hat{H}}{\partial \xi_i} |\Psi_I\rangle + \mathrm{c.c.}\right) \\
		& \equiv & H_{ij}^{(0)} + H_{ij}^{(1,2)}, \label{2ndderiv}
	\end{eqnarray}
	where the resolvent $\hat{R}$ is defined by
	\begin{eqnarray}
		\hat{R} & = & \hat{Q} (\hat{Q}\hat{H}\hat{Q}-E_I)^{-1} \hat{Q}, \label{R} \\
		\hat{Q} & = & 1 - |\Psi_I\rangle \langle\Psi_I|. \label{Q}
	\end{eqnarray}
	The reduction of Eq.~\eqref{2ndderiv0} to Eq.~\eqref{2ndderiv} is elementary and follows from expanding
	\begin{equation}
		\frac{\partial}{\partial \xi_i}\left((\hat{H}-E_I)|\Psi_I\rangle\right)=0,
	\end{equation}
	and using $E_I = \langle\Psi_I|\hat{H}|\Psi_I\rangle$ and $\hat{Q}\hat{H} = \hat{H}\hat{Q}$.
	
	The term $H_{ij}^{(0)} \equiv \langle\Psi_I| \frac{\partial^2 \hat{H}}{\partial \xi_i \partial \xi_j} |\Psi_I\rangle$ is local, since the only term of $\hat{H}$ that is neither linear nor constant in the nuclear coordinates is the nuclear-nuclear repulsion term, and its second order derivative decays like $O(1/r^3)$ with respect to the internuclear distance $r$. The second term $H_{ij}^{(1,2)}$ is, however, not necessarily local because $\hat{R}$ can be nonlocal. To further analyze the second term, we perform an energy scale separation of $\hat{R}$:
	\begin{eqnarray}
		\hat{R} & = & \hat{R}^{(1)} + \hat{R}^{(2)}, \label{energysep} \\
		\hat{R}^{(1)} & = & \hat{Q}(1-f(\hat{Q}\hat{H}\hat{Q}-E_I))(\hat{Q}\hat{H}\hat{Q}-E_I)^{-1}\hat{Q}, \label{energysep1} \\
		\hat{R}^{(2)} & = &\hat{Q}f(\hat{Q}\hat{H}\hat{Q}-E_I)(\hat{Q}\hat{H}\hat{Q}-E_I)^{-1}\hat{Q}, \label{energysep2}
	\end{eqnarray}
	where $f(x)$ is an analytic function on the whole complex plane, satisfying $f(0)=1, f'(0)=0,$ and $f(x)$ approaches 0 very quickly when $x\to\pm\infty$. For example, one can take $f(x) = \exp(-\alpha x^2)$ for some $\alpha>0$. Correspondingly, $H_{ij}^{(1,2)}$ is partitioned as
	\begin{eqnarray}
		H_{ij}^{(1,2)} & = & H_{ij}^{(1)} + H_{ij}^{(2)}, \\
		H_{ij}^{(1)} & = & \langle\Psi_I| \frac{\partial \hat{H}}{\partial \xi_j} \hat{R}^{(1)} \frac{\partial \hat{H}}{\partial \xi_i} |\Psi_I\rangle + \mathrm{c.c.}, \\
		H_{ij}^{(2)} & = & \langle\Psi_I| \frac{\partial \hat{H}}{\partial \xi_j} \hat{R}^{(2)} \frac{\partial \hat{H}}{\partial \xi_i} |\Psi_I\rangle + \mathrm{c.c.}.
	\end{eqnarray}
	
	The Taylor expansion of Eq.~\eqref{energysep1} converges quickly, since the function $(1-f(x))x^{-1}$ is analytic on the whole complex plane:
	\begin{equation}
		\hat{R}^{(1)} = \hat{Q}\sum_{n=1}^{\infty} a_n (\hat{Q}\hat{H}\hat{Q}-E_I)^n\hat{Q} = \hat{Q}\sum_{n=1}^{\infty} a_n (\hat{H}-E_I)^n\hat{Q}, \label{Taylor}
	\end{equation}
	where $a_n$ are the Taylor coefficients of the function $(1-f(x))x^{-1}$. The $n=0$ term is zero because the eigenvalue $E_I$ of $\hat{H}$ gives a zero contribution to the right hand side of Eq.~\eqref{Taylor}, due to the projector $\hat{Q}$. The convergence of the Taylor series Eq.~\eqref{Taylor} implies that only a finite number of the series is necessary for expanding $\hat{R}^{(1)}$ to a given accuracy. As $\hat{H}$ is local, it follows that powers of $(\hat{H}-E_I)$, as well as a finite sum of different powers of $(\hat{H}-E_I)$, are also local. Since
	\begin{equation}
		\hat{Q}\frac{\partial \hat{H}}{\partial \xi_j}|\Psi_I\rangle = \left(\frac{\partial \hat{V}_{\mathrm{ne}}}{\partial \xi_j} - \langle \Psi_I| \frac{\partial \hat{V}_{\mathrm{ne}}}{\partial \xi_j}|\Psi_I\rangle \right) |\Psi_I\rangle \label{QHPsi}
	\end{equation}
	is also local, in the sense that the nuclear attraction potential derivative $\frac{\partial \hat{V}_{\mathrm{ne}}}{\partial \xi_j}$ is only non-negligible near the atom corresponding to the nuclear coordinate $\xi_j$, this suggests that $H_{ij}^{(1)}$ is local. Note that here we have used the fact that $\frac{\partial \hat{H}}{\partial \xi_j} = \frac{\partial \hat{V}_{\mathrm{ne}}}{\partial \xi_j}$.
	
	We now discuss $\hat{R}^{(2)}$. Only those eigenvalues of $\hat{H}$ that are close to $E_I$ contribute non-negligibly to this term. These eigenvalues $\{E_J, J\in P\}$ (where $P$ is the set of indices of eigenvalues that are close to $E_I$) yield a low-rank contribution to $\hat{R}^{(2)}$, and therefore to the Hessian:
	\begin{eqnarray}
		\hat{R}^{(2)} & \approx & \sum_{J\in P} f(E_J-E_I)(E_J-E_I)^{-1} |\Psi_J\rangle \langle\Psi_J|, \\
		H_{ij}^{(2)} & \approx & \sum_{J\in P} f(E_J-E_I)(E_J-E_I)^{-1} \langle\Psi_I| \frac{\partial \hat{H}}{\partial \xi_j} |\Psi_J\rangle \langle\Psi_J| \frac{\partial \hat{H}}{\partial \xi_i}|\Psi_I\rangle \nonumber\\
		& = & -\sum_{J\in P} f(E_J-E_I)(E_J-E_I) d^{IJ}_j d^{JI}_i,
	\end{eqnarray}
	where
	\begin{equation}
		d^{IJ}_i = \langle \Psi_I|\frac{\partial \Psi_J}{\partial \xi_i}\rangle = (E_J-E_I)^{-1} \langle\Psi_I| \frac{\partial \hat{H}}{\partial \xi_i} |\Psi_J\rangle = -d^{JI}_i
	\end{equation}
	is the nonadiabatic coupling vector between state $I$ and state $J$\cite{NACTDDFT}.
	
	Therefore, we conclude that: (1) the Hessian of a large molecule is not necessarily local; (2) when there are other electronic states that are energetically very close to the electronic state being studied, the locality of the Hessian tends to be worse; and (3) even when the Hessian is not local, it can still be well approximated by a local component plus a low-rank component:
	\begin{equation}
		\mathbf{H} \approx \mathbf{H}^{\mathrm{local}} + \mathbf{H}^{\mathrm{lowrank}}, \label{HlHlr}
	\end{equation}
	or equivalently speaking, even when the off-diagonal blocks of the Hessian are non-negligible, they still possess low numerical rank. The last point is central to the algorithm presented herein, as it allows the Hessian to be expressed in far fewer than $O(N^2)$ parameters, even when a naive exploitation of its sparsity would not allow for such an efficient data compression.
	
	For methods where the Hellmann-Feynman theorem does not hold (e.g.~when the method uses an atom-centered basis which gives rise to the Pulay term\cite{Pulay1969}, or when the method is not variational), the above discussions are no longer rigorous. However, the qualitative conclusions should probably still hold in these cases, because all decent theoretical methods are reasonable approximations of full configuration interaction (FCI) with a complete basis set (CBS), and our assumptions are well satisfied by FCI/CBS.
	
	\section{The O1NumHess algorithm}
	
	\subsection{Recapitulation of the conventional seminumerical Hessian algorithm} \label{sec:conventional}
	
	Before we discuss how to use the ODLR property of Hessians to recover the Hessian using $O(1)$ gradients, we briefly review the conventional seminumerical Hessian algorithm which requires $O(N_{\mathrm{atom}})$ gradients.
	In conventional seminumerical Hessians, a set of gradients $g_{i}^{(j)}, i=1,\ldots,3N_{\mathrm{atom}}, j=1,\ldots,N_{\textrm{displ}}$ must be computed on a set of displaced geometries $\xi_{i}^{(j)}, i=1,\ldots,3N_{\mathrm{atom}}, j=1,\ldots,N_{\textrm{displ}}$ (for simplicity, we temporarily consider only single-sided finite difference, where the number of displacements, $N_{\textrm{disp}}$, is $3N_{\mathrm{atom}}$):
	\begin{eqnarray}
		g_{i}^{(j)} & = & g_{i}(\{\xi_{i'}^{(j)}, i'=1,\ldots,3N_{\mathrm{atom}}\}), \\
		\xi_{i}^{(j)} & = & \xi_{i}^{(0)} + \delta_{ij}\eta, \label{cart_disp}
	\end{eqnarray}
	where $\eta$ is the displacement step length, typically on the order of $0.001 \sim 0.01 \textrm{~Bohr}$. In BDF, we choose $\eta = 0.005$ Bohr as the default value, which is used in the rest of this manuscript. The elements of the Hessian are then estimated as
	\begin{equation}
		H_{ij} = \frac{g_{i}^{(j)}}{\eta}.
	\end{equation}
	
	Note that Eq.~\eqref{cart_disp} is not the only way of choosing the displacements. One can in principle choose any set of displacements $\{\xi_{i}^{(j)}\}$:
	\begin{equation}
		\xi_{i}^{(j)} = \xi_{i}^{(0)} + \Delta\xi_{i}^{(j)},
	\end{equation}
	provided that the norms of the displacements, $n_j = \sqrt{\sum_i(\Delta\xi_{i}^{(j)})^2}$, are sufficiently small. In other words, in any given displaced geometry, instead of having only one of the atoms perturbed, several or even all atoms of the molecule can be perturbed simultaneously in different directions. Then the Hessian can be obtained by inverting the relation
	\begin{equation}
		g_{i}^{(j)} = \sum_k^{3N_{\mathrm{atom}}} H_{ik}\Delta\xi_{k}^{(j)}. \label{Hdxi0}
	\end{equation}
	This can be written in the more compact, matrix form, after normalization of the gradient and displacement vectors:
	\begin{equation}
		\mathbf{g} = \mathbf{H}\mathbf{\Delta\xi}, \quad g_{ij} = g_{i}^{(j)}/n_j, \quad (\mathbf{\Delta\xi})_{ij} = \Delta\xi_{i}^{(j)}/n_j, \label{Hdxi}
	\end{equation}
	which gives
	\begin{equation}
		\mathbf{H} = \mathbf{g}\mathbf{\Delta\xi}^{-1}. \label{dxig}
	\end{equation}
	From Eq.~\eqref{dxig}, it is evident that $\mathbf{\Delta\xi}$ must have full rank (or equivalently, the set of chosen displacement directions span the space of all possible displacement directions) in order for the matrix inverse to be well-defined, and this is why the conventional numerical Hessian algorithm requires $N_{\textrm{displ}} = 3N$, even when multiple atoms are allowed to be perturbed simultaneously. Moreover, even $N_{\textrm{displ}} = 3N_{\mathrm{atom}}$ does not guarantee a full rank matrix $\mathbf{\Delta\xi}$, thus the displacements $\mathbf{\Delta\xi}$ must be chosen with some care. Some marginal savings can be further made by exploiting translational and rotational invariance, avoiding the calculation of at most 6 gradients (for details, see Appendix~\ref{appendix:transrot}).
	
	\subsection{Estimating a local Hessian using $O(1)$ gradients}
	
	We now discuss potential savings that can be made via exploiting the locality of the Hessian, in case the Hessian is local.
	A simple idea is to minimize the error of the predicted gradient, with a penalty term for the Hessian matrix elements that correspond to distant atom pairs. Specifically, the following cost function is minimized (where $\|\cdot\|$ denotes the Frobenius norm, and $\mathbf{W}\cdot\mathbf{H}$ denotes the Hadamard product):
	\begin{equation}
		\mathrm{cost}(\mathbf{H}) = \| \mathbf{g} - \mathbf{H}\mathbf{\Delta\xi} \|^2 + \lambda \| \mathbf{W}\cdot\mathbf{H} \|^2, \label{cost}
	\end{equation}
	subject to the constraint that $\mathbf{H}$ is symmetric. The penalty matrix $\mathbf{W}$ is given by
	\begin{equation}
		W_{ij} = \max(0, r_{ij} - r_i^{\mathrm{vdW}} - r_j^{\mathrm{vdW}} - \Delta r^{(1)})^\beta, \label{W}
	\end{equation}
	which depends on the distance $r_{ij}$ between the atoms corresponding to the coordinates $i$ and $j$, and the atoms' van der Waals radii ($r_i^{\mathrm{vdW}}$ and $r_j^{\mathrm{vdW}}$; taken as the UFF radii\cite{UFF}). $\Delta r^{(1)}$ is a cutoff parameter that modulates the range of the penalty: only those atom pairs that satisfy $r_{ij} > r_i^{\mathrm{vdW}} + r_j^{\mathrm{vdW}} + \Delta r^{(1)}$ are deemed sufficiently far so that the Hessian matrix elements over them receive a penalty. Empirically, the parameters $\lambda$ = 0.01 a.u., $\Delta r^{(1)}$ = 1.0 Bohr, and $\beta = 3/2$ work well for a wide range of Hessians, although we allow the user to tune $\Delta r^{(1)}$, which will control the computational cost and accuracy of the Hessian in the remaining parts of the algorithm.
	Thanks to this penalty term, we can have $N_{\textrm{disp}}<3N_{\mathrm{atom}}$ without making the problem under-determined.
	Furthermore, when $r_{ij} > r_i^{\mathrm{vdW}} + r_j^{\mathrm{vdW}} + \Delta r^{(2)}$ (where $\Delta r^{(2)} > \Delta r^{(1)}$ is another cutoff parameter), we constrain $H_{ij}$ to be exactly zero. This does not change the accuracy of the recovered Hessian noticeably, but significantly reduces the number of unknown parameters to be solved. A good compromise of accuracy and cost is given by $\Delta r^{(2)} = \Delta r^{(1)} +$ 5.0 Bohr. In the following, we will call atom pairs (or nuclear coordinate pairs) that satisfy $r_{ij} > r_i^{\mathrm{vdW}} + r_j^{\mathrm{vdW}} + \Delta r^{(2)}$ ``far-range pairs'', and those that satisfy $r_{ij} \le r_i^{\mathrm{vdW}} + r_j^{\mathrm{vdW}} + \Delta r^{(1)}$ ``near-range pairs''. Atom or nuclear coordinate pairs that are neither far-range nor near-range are called middle-range pairs.
	
	We can now solve for the Hessian elements $H_{ij}$ over near-range and middle-range pairs by differentiating Eq.~\eqref{cost} with respect to these matrix elements, yielding a set of $3N_{\mathrm{atom}}N_{\textrm{displ}}$ linear equations where $H_{ij}$ are the unknowns:
	\begin{equation}
		(\mathbf{\tilde{D}}^T\mathbf{\tilde{D}} + \mathbf{\tilde{D}}'^T\mathbf{\tilde{D}}' + 2\lambda \mathbf{\tilde{W}}\cdot\mathbf{\tilde{W}})\mathbf{\tilde{H}} = \mathbf{\tilde{D}}^T\mathbf{\tilde{g}} + \mathbf{\tilde{g}}^T\mathbf{\tilde{D}}, \label{DDH}
	\end{equation}
	where (note that $\mathbf{\tilde{g}}$ and $\mathbf{\tilde{H}}$ are vectors, while $\mathbf{\tilde{D}}$ is a matrix; $(ij)$ is a compound index where $i=1,\ldots,3N_{\mathrm{atom}}$ and $j=1,\ldots,N_{\textrm{displ}}$, while $(kl)$ and $(k'l')$ are near- or middle-range pairs)
	\begin{equation}
		\tilde{g}_{(ij)} = g_{ij},
	\end{equation}
	\begin{equation}
		\tilde{D}_{(ij)(kl)} = \delta_{il}(\mathbf{\Delta\xi})_{kj},
	\end{equation}
	\begin{equation}
		\tilde{D}_{(ij)(kl)}' = \delta_{ik}(\mathbf{\Delta\xi})_{lj},
	\end{equation}
	\begin{equation}
		\tilde{W}_{(kl)(k'l')} = \delta_{kk'}\delta_{ll'}W_{kl},
	\end{equation}
	\begin{equation}
		\tilde{H}_{(kl)} = H_{kl}.
	\end{equation}
	Eq.~\eqref{DDH} is a very sparse linear system, with dimensions $N_{\textrm{nz}} \times N_{\textrm{nz}}$, where $N_{\textrm{nz}}$ is the number of near-range and middle-range pairs $(ij), i\le j$.
	Like all sparse linear systems (such as the CP-SCF equations), Eq.~\eqref{DDH} can be solved by e.g.~the generalized minimal residual (GMRES) method\cite{GMRES}, where the sparse matrices $\mathbf{\tilde{D}}^\dag\mathbf{\tilde{D}}$ and $\mathbf{\tilde{D}}'^\dag\mathbf{\tilde{D}}'$ are never stored in full, but its products with vectors are evaluated on the fly. The matrices $\mathbf{\tilde{D}}^\dag\mathbf{\tilde{D}}$ and $\mathbf{\tilde{D}}'^\dag\mathbf{\tilde{D}}'$ in Eq.~\eqref{DDH} are symmetric and non-negative definite, which is an extra bonus since the conjugate gradient method, which is simpler to implement and less memory-intensive, may be used in place of GMRES when necessary.
	
	If $N_{\textrm{displ}}$ is chosen such that $3N_{\mathrm{atom}}N_{\textrm{displ}} \ge N_{\textrm{nz}}$ (which means that $N_{\textrm{displ}}$ can be asymptotically $O(1)$, since $N_{\textrm{nz}}$ is $O(N_{\mathrm{atom}})$), and the displacements $\mathbf{\Delta\xi}$ are suitably chosen such that $\mathbf{\tilde{D}}$ has full rank, it follows that the non-zero elements of $\mathbf{H}$ can be uniquely determined, since the left hand side matrix of Eq.~\eqref{DDH} is positive definite. Furthermore, our exploitation of the symmetry of $\mathbf{H}$ reduces $N_{\textrm{nz}}$ by almost one half, and is thus expected to even further reduce the minimum $N_{\textrm{displ}}$ required to make the solution of the non-zero Hessian elements unique.
	
	\subsection{Estimating an ODLR Hessian using $O(1)$ gradients}
	
	The above algorithm always gives a local Hessian (hereafter denoted as $\mathbf{\bar{H}}^{\mathrm{local}}$), but as we have already seen in Section \ref{sec:ODLR}, this is not always guaranteed. When the exact Hessian is not local but still satisfies the ODLR property (Eq.~\eqref{HlHlr}), $\mathbf{\bar{H}}^{\mathrm{local}}$ will be a poor approximation to the Hessian $\mathbf{H}$ (as the predicted Hessian gives zero matrix elements for the far-range pairs). However, $\mathbf{\bar{H}}^{\mathrm{local}}$ should still be a reasonable approximation of $\mathbf{H}^{\mathrm{local}}$, and therefore $\mathbf{H} - \mathbf{\bar{H}}^{\mathrm{local}} \approx \mathbf{H}^{\mathrm{lowrank}}$ is expected to have low numerical rank (Eq.~\eqref{HlHlr}). Therefore, we estimate the Hessian $\mathbf{H}$ by repeatedly finding a low-rank (but not necessarily symmetric) correction that makes the Hessian reproduce the gradients $\mathbf{g}$, followed by symmetrization:
	\begin{enumerate}
		\item Set $\mathbf{H}_0 = \mathbf{\bar{H}}^{\mathrm{local}}$. Set $n=0$.
		\item Update the Hessian $\mathbf{H}_n$ by adding a low-rank correction, so that the updated Hessian exactly satisfies Eq.~\eqref{Hdxi}, but may be unsymmetric: \label{LR_startiter}
		\begin{equation}
			\mathbf{H}_{n+1}^{\mathrm{unsym}} = \mathbf{H}_n + (\mathbf{g}^{\mathrm{scaled}}-\mathbf{H}_n\mathbf{\Delta\xi}^{\mathrm{scaled}})(\mathbf{\Delta\xi}^{\mathrm{scaled}})^T.
		\end{equation}
		Here, we scale down a gradient (and its corresponding displacement direction) when the gradient's norm is large. This makes the low-rank correction prioritize on improving the low frequency modes of the Hessian, which are more important than high frequency modes for accurate entropies and Gibbs free energies:
		\begin{equation}
			g^{\mathrm{scaled}}_{ij} = g_{ij}\frac{\epsilon}{\max(\epsilon,\sqrt{\sum_i(g_{ij})^2})},
		\end{equation}
		\begin{equation}
			(\mathbf{\Delta\xi}^{\mathrm{scaled}})_{ij} = (\mathbf{\Delta\xi})_{ij}\frac{\epsilon}{\max(\epsilon,\sqrt{\sum_i(g_{ij})^2})},
		\end{equation}
		where $\epsilon$ is chosen to be $10^{-3}$.
		\item Symmetrize the Hessian, at the expense of making the Hessian not satisfying Eq.~\eqref{Hdxi}:
		\begin{equation}
			\mathbf{H}_{n+1} = \frac{1}{2}(\mathbf{H}_{n+1}^{\mathrm{unsym}}+(\mathbf{H}_{n+1}^{\mathrm{unsym}})^T).
		\end{equation}
		\item If self-consistency is achieved such that the relative change of $\|\mathbf{g}^{\mathrm{scaled}}-\mathbf{H}_n\mathbf{\Delta\xi}^{\mathrm{scaled}}\|$ during one iteration is within $10^{-8}$, or if $\|\mathbf{g}^{\mathrm{scaled}}-\mathbf{H}_n\mathbf{\Delta\xi}^{\mathrm{scaled}}\| < 10^{-8}$, terminate the procedure and return $\mathbf{H} = \mathbf{H}_{n+1}$. Otherwise increment $n$ and go to step \ref{LR_startiter}.
	\end{enumerate}
	In essence, while it is difficult to design a correction that is low-rank, symmetric, and consistent with the calculated gradients, the algorithm alternates between (1) adding a low-rank correction to the Hessian so that it exactly reproduces the calculated gradients $\mathbf{g}$ given $\mathbf{\Delta\xi}$, and (2) symmetrizing the Hessian. Although the exact Hessian of a system is always symmetric, the use of a finite step length in the numerical differentiations may result in gradients that cannot be reproduced by any symmetric Hessian. The norm of $\mathbf{g}^{\mathrm{scaled}}-\mathbf{H}\mathbf{\Delta\xi}^{\mathrm{scaled}}$ is thus not necessarily zero at convergence, and correlates with the numerical error of the computed Hessian.
	
	\subsection{Generation of the displacement directions}
	
	Now it only remains to specify the method of generating the displacement directions $\mathbf{\Delta\xi}$. To see what properties $\mathbf{\Delta\xi}$ has to satisfy to deliver accurate Hessians, we first consider a system composed of $N_{\mathrm{frag}}$ infinitely separated molecular fragments $\{M_k, k=1,\ldots,N_{\mathrm{frag}}\}$, where all fragments are small enough such that all intra-fragment atom pairs are short-ranged. In this case, the local part of the Hessian $\mathbf{\bar{H}}^{\mathrm{local}}$ is block-diagonal, with one block per fragment. We now consider the projection of all displacement directions $\mathbf{\Delta\xi}$ onto a given fragment $M_k$, $\mathbf{\Delta\xi}^{M_k}$:
	\begin{equation}
		(\mathbf{\Delta\xi}^{M_k})_{ij} \equiv (\mathbf{\Delta\xi})_{i^{(M_k)}j}, \quad i=1,\ldots,3N_{M_k},
	\end{equation}
	where $N_{M_k}$ is the number of atoms of fragment $M_k$, and $i^{(M_k)}$ is the index of fragment $M_k$'s $i$-th nuclear coordinate in the whole molecular system; $(\mathbf{\Delta\xi}^{M_k})_{ij}$ is the $i$-th component of the $j$-th projected displacement direction.
	It is clear that $\mathbf{\Delta\xi}^{M_k}$ must constitute a complete (or overcomplete) set of displacements for fragment $M_k$, so that the gradients along these displacement directions provide enough information to determine the respective block of the Hessian. This is like how we need the chosen displacement directions to span the space of all possible displacement directions of the whole molecule, if we do not assume any structure (like locality) of the whole molecule's Hessian (Section~\ref{sec:conventional}).
	
	Even for systems that are not composed of small, non-interacting fragments, one would still hope that the displacement directions form a ``locally (over)complete'' set around every atom. Thus, for each atom $A$, we project the displacement directions $\mathbf{\Delta\xi}$ onto the neighborhood $\mathcal{N}(A)$ of atom $A$, defined as the set of atoms that form near-range pairs with $A$:
	\begin{equation}
		(\mathbf{\Delta\xi}^{\mathcal{N}(A)})_{ij} \equiv (\mathbf{\Delta\xi})_{i^{\mathcal{N}(A)}j}.
	\end{equation}
	We should then require $\mathbf{\Delta\xi}^{\mathcal{N}(A)}$ to be an (over)complete set of basis vectors for expanding all possible displacement directions of the atoms in $\mathcal{N}(A)$. Therefore, if we have already generated $N_{\mathrm{displ}}$ displacement directions, the projection of the $(N_{\mathrm{displ}}+1)$-th displacement direction onto $\mathcal{N}(A)$ should be approximately orthogonal with, or at least linearly independent of, all the previous displacement directions, to the extent possible. Namely (as before, $N_{\mathcal{N}(A)}$ is the number of atoms in $\mathcal{N}(A)$):
	\begin{equation}
		\sum_{i}^{3N_{\mathcal{N}(A)}} (\mathbf{\Delta\xi}^{\mathcal{N}(A)})_{i(N_{\mathrm{displ}}+1)}(\mathbf{\Delta\xi}^{\mathcal{N}(A)})_{ij} \approx 0, \quad \forall j=1,\ldots,N_{\mathrm{displ}}.
	\end{equation}
	
	Inspired by the above observations, we generate the displacement directions using the following approach:
	\begin{enumerate}
		\item Calculate a cheap model Hessian $\mathbf{H}^{\mathrm{Swart}}$ of the system, using a modified version of Swart et al.'s method\cite{Swart}. Details are given in Appendix~\ref{appendix:Swart}.
		\item Generate $N_{\mathrm{displ}}=7$ displacement directions ($N_{\mathrm{displ}}=6$ for linear molecules; in the following we will assume that the molecule is non-linear), consisting of 3 translations, 3 rotations and the symmetric breathing mode (i.e.~the vibrational mode where all interatomic distances expand and contract by the same ratio). The importance of adding the symmetric breathing mode to the list of displacement directions will be revealed later.
		\item Loop until no new displacement directions are generated:
		\begin{enumerate}
			\item For each atom $A$, take the subblock of $\mathbf{H}^{\mathrm{Swart}}$ over $\mathcal{N}(A)$, project out the space spanned by the existing $N_{\mathrm{displ}}$ displacement directions $\mathbf{\Delta\xi}^{\mathcal{N}(A)}$, and diagonalize the resulting matrix. Set the ``raw local mode'' $(\mathbf{\Delta\xi}^{\mathcal{N}(A),\mathrm{local}})_{i(N_{\mathrm{displ}}+1)}, i=1,\ldots,3N_{\mathcal{N}(A)}$ as the eigenvector with the largest eigenvalue. This is equivalent to choosing a vibrational mode that is as stiff as possible, among the vectors that are orthogonal to all vectors in $\mathbf{\Delta\xi}^{\mathcal{N}(A)}$. Such a choice is empirically found to work well. If $N_{\mathrm{displ}}$ is equal to or greater than $3N_{\mathcal{N}(A)}$, the matrix to be diagonalized will be a zero matrix, in which case we set $(\mathbf{\Delta\xi}^{\mathcal{N}(A),\mathrm{local}})_{i(N_{\mathrm{displ}}+1)} = 0, \forall i$.
			\item Sum up the raw local modes of each atom, multiplied by appropriate sign factors $\sigma_A^{(N_{\mathrm{displ}}+1)} = \pm 1$:
			\begin{equation}
				\mathbf{\Delta\xi}^{\mathrm{global},(N_{\mathrm{displ}}+1)}_i = \sum_{A \mathrm{~s.t.~} i\in\mathcal{N}(A)} \sigma_A^{(N_{\mathrm{displ}}+1)} (\mathbf{\Delta\xi}^{\mathcal{N}(A),\mathrm{local}})_{i(N_{\mathrm{displ}}+1)}
			\end{equation}
			The sign factors are chosen to maximize the norm of $\mathbf{\Delta\xi}^{\mathrm{global}}$, via a greedy algorithm (i.e.~the sign of the $A$-th term is chosen while keeping the signs of the previous $A-1$ terms fixed, and neglecting the terms after the $A$-th term). In other words, we choose the signs so that the raw local modes of different atoms form in-phase combinations with each other whenever possible. When this does not suffice to determine a sign factor $\sigma_A^{(N_{\mathrm{displ}}+1)}$ (for example, because $\mathcal{N}(A)$ does not overlap with the neighborhoods of any previous atoms, $\mathcal{N}(B), B<A$), we choose $\sigma_A^{(N_{\mathrm{displ}}+1)}$ to be the sign of the element of $(\mathbf{\Delta\xi}^{\mathcal{N}(A),\mathrm{local}})_{i(N_{\mathrm{displ}}+1)}, i=1,\ldots,3N_{\mathcal{N}(A)}$ with the largest absolute value. This does not necessarily improve the results, but makes our method deterministic (except for the unlikely case where two elements of $\mathbf{\Delta\xi}^{\mathcal{N}(A),\mathrm{local}}$ both have the largest absolute value, but have different signs).
			\item If $\mathbf{\Delta\xi}^{\mathrm{global},(N_{\mathrm{displ}}+1)} = \mathbf{0}$ (which can only happen when $(\mathbf{\Delta\xi}^{\mathcal{N}(A),\mathrm{local}})_{i(N_{\mathrm{displ}}+1)}=0, \forall A,i$), exit the loop and terminate the procedure.
			\item Otherwise, orthonormalize $\mathbf{\Delta\xi}^{\mathrm{global},(N_{\mathrm{displ}}+1)}$ against all existing displacement directions, scale it so that the element with the largest absolute value becomes the step length  $\eta$, and set the $(N_{\mathrm{displ}}+1)$-th displacement direction as the resulting vector. Increment $N_{\mathrm{displ}}$ and go back to step (a).
		\end{enumerate}
	\end{enumerate}
	
	By this way, we arrive at a set of displacement directions with the following properties:
	\begin{enumerate}
		\item Unless the molecule is in an external field that breaks its translational and/or rotational symmetry, the gradients along the first 6 displacement directions do not need to be calculated. The gradients along the translational modes are always zero, while the gradients along the rotational modes can be inferred from the gradient at the equilibrium geometry (Appendix~\ref{appendix:transrot}).
		\item The leading cubic anharmonicity is largely captured by the seventh displacement direction (the symmetric breathing mode), so that once the seventh displacement direction is treated by double-sided numerical differentiation, the subsequent displacement directions can be treated by single-sided numerical differentiation with little loss of accuracy. This is because (a) bond stretching vibrations usually have larger diagonal anharmonic constants compared to other vibration modes (see e.g.~Table 2 of Ref.~\cite{tetrafluoroethylene_anharmonic}, Table 2 of Ref.~\cite{formic_acid_anharmonic} and Table 6 of Ref.~\cite{furan_anharmonic}, where the diagonal anharmonic constants of bond stretching vibrations are 1-2 orders of magnitudes larger than those of other modes); (b) the diagonal anharmonic constants of most bonds are negative, such that elongating a bond reduces the bond's force constant. Therefore, in the symmetric breathing mode, the diagonal anharmonic constants of all bonds add up. All the subsequent displacement directions have been orthogonalized against the symmetric breathing mode, and therefore have roughly equal contributions from bond stretching and bond compression; the diagonal anharmonic constants thus tend to cancel out.
		\item Most importantly, as each atom has only $O(1)$ neighbors, the loop that generates displacement directions terminates after $O(1)$ steps due to $(\mathbf{\Delta\xi}^{\mathcal{N}(A),\mathrm{local}})_{i(N_{\mathrm{displ}}+1)}=0, \forall A,i$.
	\end{enumerate}
	
	Finally, when the resulting Hessian has negative eigenvalues, we orthogonalize the corresponding eigenvectors against existing displacement directions, and add them to the list of displacement directions (if this would result in an over-complete set of displacement directions, we add just enough eigenvectors to make the set complete). We then calculate the gradients along these directions as well, and estimate the Hessian again using the O1NumHess algorithm, using the full set of gradients and displacement directions. This helps to minimize the number of ``false imaginary frequencies'' due to numerical error, and improves the accuracy of entropies and Gibbs free energies. Although it is in principle possible that the number of negative Hessian eigenvalues increases with system size, we can always select only those negative Hessian eigenvalues that are closest to zero (since they are the most likely to be a result of numerical error), and perform the aforementioned steps only on this subset of eigenvectors. Therefore, the number of gradients required by O1NumHess can remain $O(1)$ in spite of this extra step.
	
	\section{Implementation}
	
	We implemented our algorithm in two open-source Python 3 libraries, \verb|O1NumHess| (https://github.com/ilcpm/O1NumHess) and \verb|O1NumHess_QC| \\
	(https://github.com/ilcpm/O1NumHess\_QC), both available on GitHub. \verb|O1NumHess| is intended to be a general tool for calculating the Hessian with $O(1)$ gradients, and can in principle be applied to problems irrelevant to quantum chemistry; for example, it does not assume that the number of ``nuclear coordinates'' is a multiple of three. Accordingly, the user has to provide the distances between every pair of ``nuclear coordinates'' $r_{ij}$, the model Hessian, and information on whether certain gradients are known in advance and do not need to be calculated, etc. Users interested in using the present algorithm for general (i.e.~not necessarily related to computational chemistry) Hessian estimation problems are encouraged to adapt the \verb|O1NumHess| library to their needs. In principle, the algorithm can be applied to any problem where a ``distance'' can be defined between two arbitrary variables $\xi_i$ and $\xi_j$ of the multivariate function $E(\xi_1,\ldots,\xi_n)$ whose Hessian is to be sought for, and blocks of the Hessian corresponding to ``far'' groups of variables have low rank. The accuracy (and cost) of the calculation can be controlled by a single parameter, $\Delta r^{(1)}$; the other parameters do not significantly influence the results for nuclear Hessians, although we encourage the users to experiment over these parameters for Hessians other than the nuclear Hessian. The traditional single-sided and double-sided differentiation algorithms were also implemented in \verb|O1NumHess| for comparison purposes.
	
	The \verb|O1NumHess_QC| library relies on \verb|O1NumHess| as its core subroutine, and specializes for Hessian calculations in the field of quantum chemistry. It provides the distance matrix, model Hessian etc.~necessary for the \verb|O1NumHess| library. Meanwhile, \verb|O1NumHess_QC| also implements interfaces that call electronic structure programs and parse the resulting gradients. The \verb|O1NumHess_QC| library was interfaced with the BDF program\cite{BDF1,BDF2,BDF3,BDF4,BDF5} in a two-way fashion, such that the user can call the \verb|O1NumHess_QC| library by adding the \verb|O1NumHess| keyword to the \verb|$bdfopt| block of the input file, and \verb|O1NumHess_QC| in turn calls BDF to calculate the required gradients. We also interfaced \verb|O1NumHess_QC| with ORCA\cite{ORCA1,ORCA2,ORCA3,ORCA4,ORCA5}, though in this case one has to calculate the Hessian by directly calling the Python functions in the \verb|O1NumHess_QC| library.
	
	During the calculation, the gradient at the equilibrium geometry is first computed using all available processors. The gradients at the displaced geometries are then calculated in an embarrassingly parallel fashion, using the wavefunction at the equilibrium geometry as the initial guess. By default, each gradient calculation is performed serially, but the user can also ask the program to calculate each gradient in parallel using a fraction of the total number of available processors, which helps to improve load balance (by increasing the number of task batches) at the expense of reducing the parallel efficiency of each gradient calculation itself. When there is an insufficient number of gradient calculations left, the program increases the number of processors used by each remaining gradient calculation, to better exploit the available computational resources.
	
	\section{Numerical benchmarks}
	
	All benchmark results were obtained using a development version of BDF as the electronic structure software, at the B3LYP-D3/def2-SV(P) level of theory. The Git commits 775fb47 and 6ee65eb of \verb|O1NumHess| and \verb|O1NumHess_QC|, respectively, were used in the benchmark calculations. For the S30L-CI set, solvation effects were added using the IEFPCM model\cite{IEFPCM1,IEFPCM2,IEFPCM3,IEFPCM4,IEFPCM5}, with chloroform as solvent. Vibrational frequency errors were computed by sorting the exact and approximate frequencies in ascending order, followed by subtraction; no attempt was made to assign the frequencies based on vibrational mode compositions or irreducible representations, which may lead to underestimations of the true frequency errors. All Gibbs free energies were calculated using the quasi rigid rotor harmonic oscillator (QRRHO)\cite{QRRHO} approach. Timing and memory usage data were obtained using 64 OpenMP threads in total, a minimum of 4 OpenMP threads per gradient calculation, on a node equipped with 64 Hygon C86 7285H processors and 256 GB of memory. 200 GB of heap memory was allocated for every analytic Hessian calculation (via the \verb|maxmem| keyword in the \verb|$resp| block of the BDF input file), except for three cases (``complex4'' of the WCCR10 set, as well as ``10AB'' and ``12AB'' of the S30L-CI set) where we had to reduce the heap memory budget to 128 GB due to increased stack memory usage. The complexes in S30L-CI whose analytic frequency calculations did not finish within 10 days (number 8, 13, 14, 20, 24), together with their constituent fragments, are not included in the analyses.
	
	\subsection{Model systems: $n$-\ce{C32H66} and \ce{C32H34}}
	
	As a first example, we apply our algorithm to the aforementioned linear molecules $n$-\ce{C32H66} and \ce{C32H34}. To exclude the error due to using a finite numerical differentiation step $\eta$, we first perform benchmark studies where the gradients $\mathbf{g}$ are calculated by multiplying the known exact Hessian of the system $\mathbf{H}$ with $\mathbf{\Delta\xi}$ (Eq.~\eqref{Hdxi}). Here ``exact Hessians'' refers to analytic Hessians, except for $S_1$ states for which it refers to seminumerical Hessians obtained from conventional double-sided differentiation. Although the exact Hessian is of course not known in a real O1NumHess calculation, the present approach simulates the results that would be obtained if $\eta$ is infinitesimal and the gradients are free of numerical error. The effect of using a finite $\eta$ will be discussed at the end of this subsection.
	
	To begin with, we study the vibrational frequency and Gibbs free energy errors as a function of $\Delta r^{(1)}$ (Table~\ref{table1}). The number of gradient evaluations (which is equal to the number of displacement directions plus two, since one has to evaluate the gradient at the equilibrium geometry, and for the symmetric breathing mode two gradients are required) increase with $\Delta r^{(1)}$, except for \ce{C32H34} ($S_1$) when going from $\Delta r^{(1)} = 0$ Bohr to $\Delta r^{(1)} = 1$ Bohr; the latter is because the Hessian has fewer imaginary frequencies with $\Delta r^{(1)} = 1$ Bohr than with $\Delta r^{(1)} = 0$ Bohr, so that fewer additional gradient evaluations are needed for improving the accuracy of imaginary modes. Nevertheless, even for the largest $\Delta r^{(1)}$ tested here, O1NumHess is still much cheaper than the conventional numerical Hessian algorithm, which can be in part attributed to the enhanced Hessian locality due to the one-dimensional character of the molecules.
	For the saturated alkane $n$-\ce{C32H66} (representative of systems without low-lying excited states and therefore with a rather local Hessian), the frequency and Gibbs free energy errors decrease steadily with the increase of $\Delta r^{(1)}$, and excellent accuracy is already obtained with $\Delta r^{(1)} = 1$ Bohr. By comparison, the frequency errors of \ce{C32H34} are much larger than those of $n$-\ce{C32H66}, though still well within the errors of typical density functionals (a few tens of cm$^{-1}$). Unlike the case of $n$-\ce{C32H66}, for the three electronic states of \ce{C32H34}, increasing $\Delta r^{(1)}$ does not lead to a consistent accuracy improvement, since increasing $\Delta r^{(1)}$ only increases the bandwidth of the local part of the Hessian $\mathbf{H}^{\mathrm{local}}$ (cf.~Eq.~\eqref{W}), but is not expected to improve the long-ranged, low-rank part $\mathbf{H}^{\mathrm{lowrank}}$.
	Based on the above results, we choose $\Delta r^{(1)} = 1$ Bohr for all subsequent tests, as a compromise between accuracy and cost (although this gives the worst frequencies for the three states of \ce{C32H34}, it yields the most accurate Gibbs free energy differences between these three states).
	
	\begin{table}
		\centering
		\caption{Errors of frequencies (cm$^{-1}$) and Gibbs free energies (kcal/mol) calculated from the O1NumHess-approximated B3LYP-D3/def2-SV(P) Hessians of ground state $n$-\ce{C32H66} and the $S_0$, $S_1$ and $T_1$ states of \ce{C32H34}, as a function of the parameter $\Delta r^{(1)}$ (Bohr), compared to those from the respective exact Hessians. The numbers of gradient evaluations required by O1NumHess is compared with those required by the conventional double-sided numerical differentiation algorithm ($6N_{\mathrm{atom}}$), and their ratios are shown in parentheses. The gradient calculations are ``simulated'' by multiplying the exact Hessian with the displacement directions; see the main text for details. MAD: mean absolute deviation. MD: mean deviation. MaxD: maximum deviation.}
		\begin{tabular}{cccccc}
			\hline\hline
			& $\Delta r^{(1)}$ & $n$-\ce{C32H66} ($S_0$) & \ce{C32H34} ($S_0$) & \ce{C32H34} ($T_1$) & \ce{C32H34} ($S_1$) \\
			\hline
			$6N_{\mathrm{atom}}$ & & 588 & 396 & 396 & 396 \\
			\hline
			& 0.0 & 42 (7\%) & 40 (10\%) & 41 (10\%) & 39 (10\%) \\
			Number of gradients & 1.0 & 53 (9\%) & 40 (10\%) & 42 (11\%) & 38 (10\%) \\
			& 2.0 & 66 (11\%) & 45 (11\%) & 46 (12\%) & 44 (11\%) \\
			\hline
			& 0.0 & 1.97 & 4.48 & 8.40 & 4.35 \\
			MAD (frequencies) & 1.0 & 0.78 & 6.88 & 11.9 & 6.06 \\
			& 2.0 & 0.30 & 6.15 & 9.64 & 3.25 \\
			\hline
			& 0.0 & 1.27 & -1.39 & -0.65 & 0.38 \\
			MD (frequencies) & 1.0 & 0.20 & 1.02 & 7.32 & 4.43 \\
			& 2.0 & -0.06 & 4.21 & 5.10 & 1.78 \\
			\hline
			& 0.0 & 15.6 & 25.7 & 81.0 & 31.4 \\
			MaxD (frequencies) & 1.0 & 8.49 & 68.9 & 94.1 & 33.4 \\
			& 2.0 & 6.39 & 50.8 & 60.9 & 22.7 \\
			\hline
			& 0.0 & 2.94 & 1.06 & 2.94 & 2.35 \\
			Error of Gibbs free energy & 1.0 & -0.43 & 2.02 & 2.71 & 2.44\\
			& 2.0 & -0.24 & 2.14 & 2.57 & 1.19 \\
			\hline\hline
		\end{tabular}\label{table1}
	\end{table}
	
\begin{figure}[!]
		\centering
		\includegraphics[width=\textwidth]{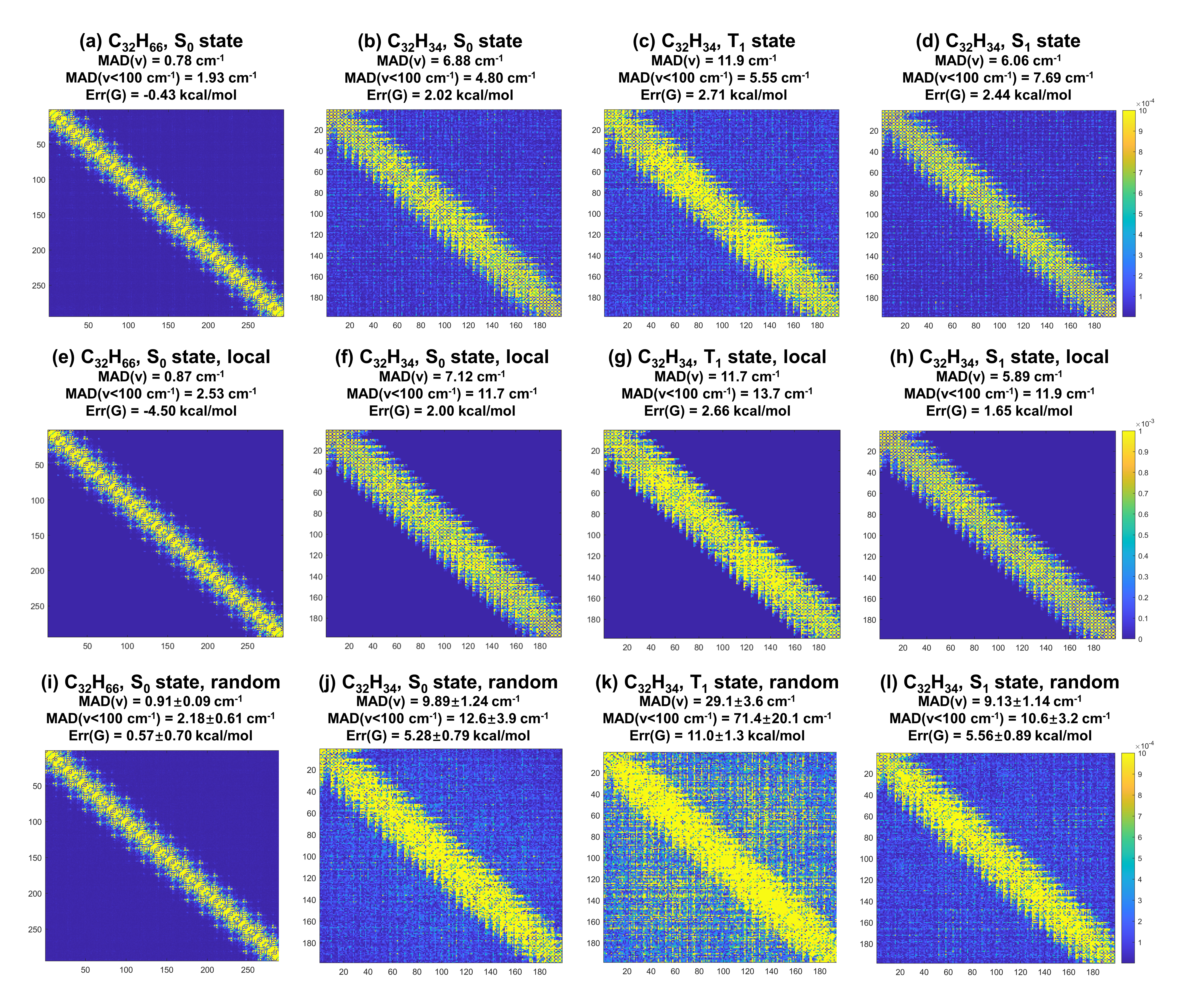}
		\caption{Heatmaps of the O1NumHess-approximated B3LYP-D3/def2-SV(P) Hessians of (a) $n$-\ce{C32H66}, as well as (b-d) the $S_0$, $S_1$ and $T_1$ states of \ce{C32H34}; (e-h) Same as (a-d) but without the low-rank correction $\mathbf{H}^{\mathrm{lowrank}}$; (i-l) Same as (a-d) but with random orthonormal displacement directions.  The matrix elements are color-coded based on their absolute values. The gradient calculations are ``simulated'' by multiplying the exact Hessian with the displacement directions; see the main text for details. Frequency MADs and Gibbs free energy errors are shown; for every calculation that uses random displacement directions, the average as well as the standard deviation of 100 runs are reported, and the heatmap from a representative run is shown.}
		\label{heatmap-o1nh}
	\end{figure}

	To gain deeper insight into the role of different components of the O1NumHess algorithm, we plotted the Hessians of the aforementioned four systems obtained with O1NumHess (Figure~\ref{heatmap-o1nh}), which can be compared with the plots of the exact Hessians in Figure~\ref{heatmap}(a-d). Adding $\mathbf{H}^{\mathrm{lowrank}}$ not only generates a non-local off-diagonal part of the Hessian, but nicely reproduces the positions of all the non-negligible off-diagonal Hessian matrix elements (Figure~\ref{heatmap-o1nh}(a-d)). Obviously, the low-rank nature of the correction plays an important role in reproducing the regular, grid-like pattern in the off-diagonal parts of the Hessian, because such a pattern inherently has low numerical rank. By comparison, $\bar{\mathbf{H}}^{\mathrm{local}}$ completely fails to recover the off-diagonal patterns of the Hessians, and predicts zero matrix elements in these regions (Figure~\ref{heatmap-o1nh}(e-h)). Although this hardly affects the MAD of all frequencies, it raises the errors of low frequency modes significantly (almost tripling the MAD for \ce{C32H34} ($S_0$)), and increases the Gibbs free energy error of $n$-\ce{C32H66} by an order of magnitude. We therefore conclude that although $\bar{\mathbf{H}}^{\mathrm{local}}$ already gives qualitatively correct frequencies, adding $\mathbf{H}^{\mathrm{lowrank}}$ significantly improves their accuracy at negligible cost.
	
	Notably, the excellent reproduction of the Hessian is also partly attributed to our judicious choice of the displacement directions. Using Schmidt-orthogonalized sets of random vectors in place of our displacement directions $\mathbf{\Delta\xi}$, we obtained up to 200\% higher MADs for all frequencies, and up to 13 times higher MADs for frequencies below 100 cm$^{-1}$, probably owing to an overestimation of the magnitudes of off-diagonal Hessian entries (as is especially evident for \ce{C32H34} ($T_1$)). The Gibbs free energies show a large spread due to stochastic noise even for $n$-\ce{C32H66}, and for the three states of \ce{C32H34} we observe huge, system-dependent systematic errors on the order of 5-10 kcal/mol, which makes the method completely unsuitable for obtaining Gibbs free energy differences, at least with the default $\Delta r^{(1)}$. In all cases, our deterministic algorithm gives lower errors than the average error of the stochastic algorithm over 100 runs, and for the free energies of the three states of \ce{C32H34} as well as the frequencies of \ce{C32H34} ($T_1$), our improvements with respect to the stochastic algorithm even exceed 3 times of the respective standard deviations. This highlights the importance of using deterministic algorithms to generate displacement directions instead of stochastic ones, which not only encourages error cancellation upon taking free energy differences, but also reduces the frequency and free energy errors themselves.
	
\begin{figure}[!]
		\centering
		\includegraphics[width=\textwidth]{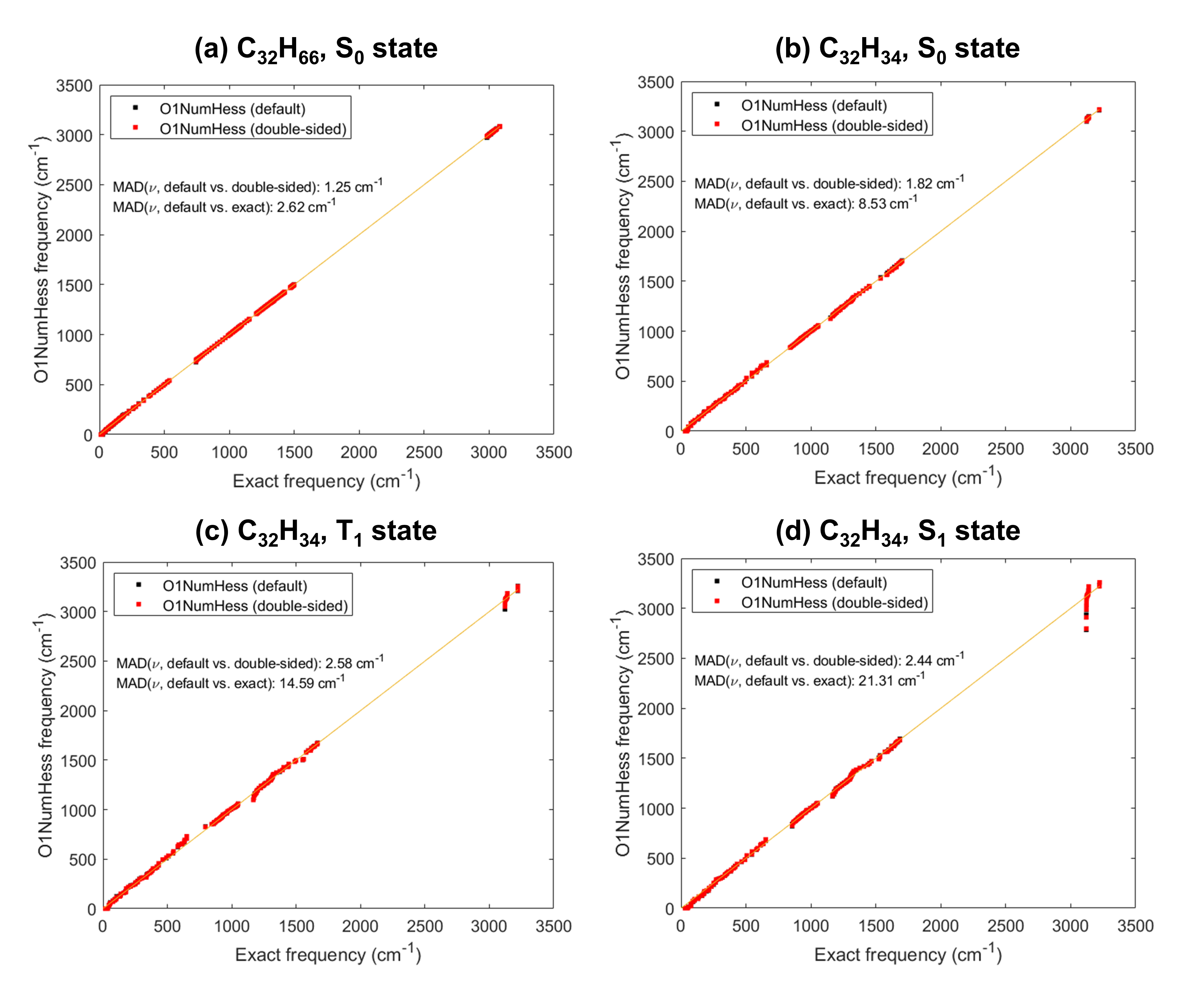}
		\caption{Frequency errors of the O1NumHess-approximated B3LYP-D3/def2-SV(P) Hessians of (a) $n$-\ce{C32H66}, as well as (b-d) the $S_0$, $S_1$ and $T_1$ states of \ce{C32H34}, with default settings (``default'') as well as with double-sided differentiation along all displacement directions (``double-sided''), compared to the frequencies of the exact Hessians.}
		\label{freq}
	\end{figure}

	Finally, we replace the ``simulated'' gradient calculations with real gradient calculations on displaced geometries, and plot the vibrational frequencies of the O1NumHess Hessians against those of the exact Hessians (Figure~\ref{freq}). Except for the $S_1$ state of \ce{C32H34}, all systems show an increase of frequency MADs (default O1NumHess frequencies versus exact frequencies) by only about 2 cm$^{-1}$ compared to the ``simulated'' gradients (Table~\ref{table1}), which is comparable to the frequency change upon replacing all single-sided numerical differentiations in the O1NumHess algorithm by double-sided ones. The abnormally large frequency error of \ce{C32H34} ($S_1$) may be attributed to the larger numerical noise of TDDFT gradients compared to ground-state gradients, which may have been further magnified by the O1NumHess algorithm to yield less accurate Hessians than when the numerical noise is absent. Overall, we conclude that our treatment of only one displacement direction using double-sided differentiation is sufficiently accurate compared to the other error sources in the algorithm.
	
	\subsection{Large covalent complexes: the WCCR10 set}
	
\begin{figure}[!]
		\centering
		\includegraphics[width=\textwidth]{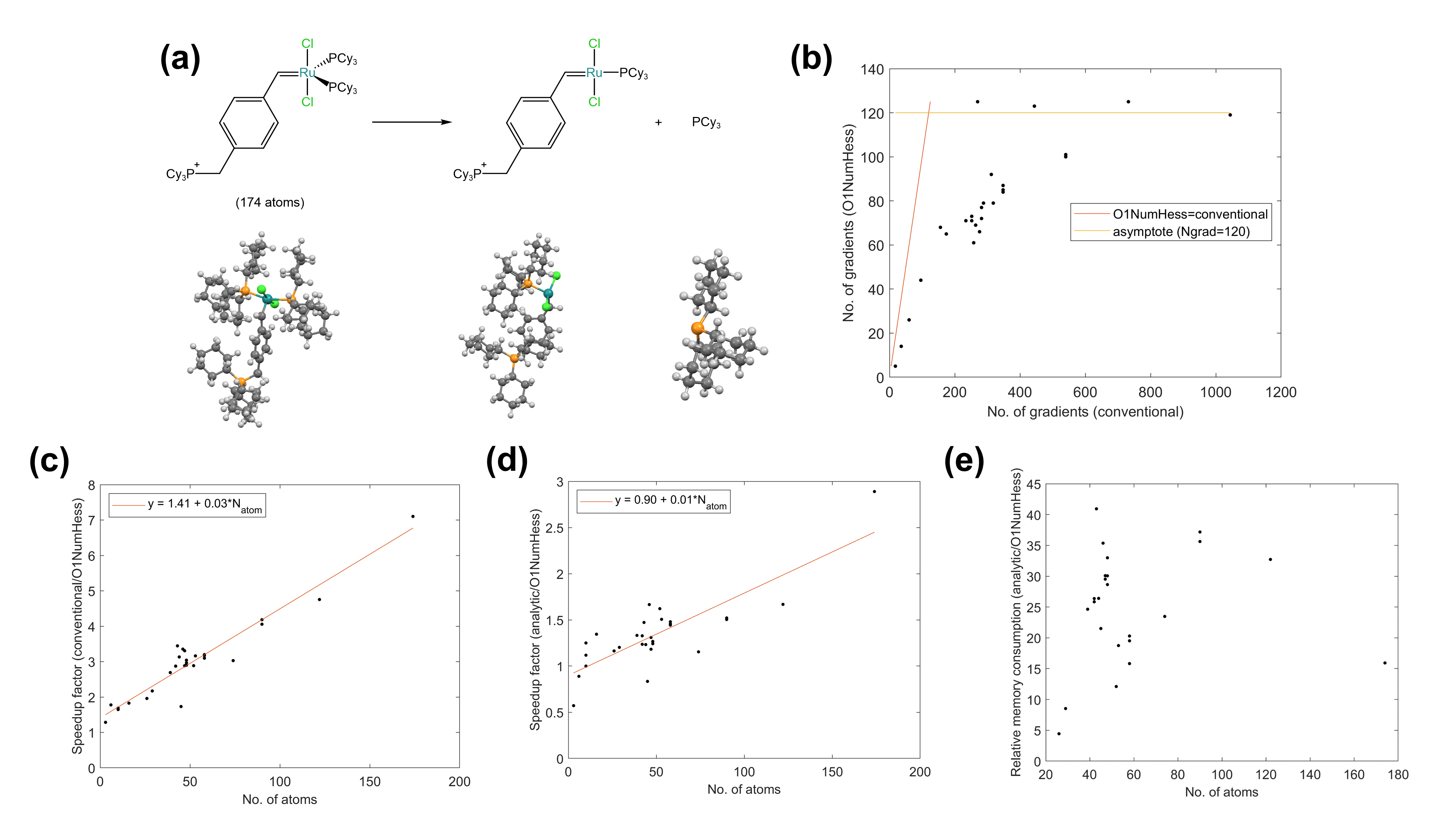}
		\caption{Efficiency of O1NumHess on the WCCR10 set. (a) A representative reaction in the WCCR10 set; (b) numbers of gradients required by O1NumHess, compared to those required by the conventional double-sided seminumerical Hessian algorithm; (c) speedup factors of O1NumHess compared to conventional double-sided seminumerical Hessians; (d) speedup factors of O1NumHess compared to analytic Hessians; (e) relative memory usage of analytic Hessian compared to O1NumHess.}
		\label{WCCR10-time}
	\end{figure}

	We then turn our attention to more realistic systems. WCCR10\cite{WCCR10} is a benchmark set for the gas phase ligand dissociation energies of transition metal complexes with tens to over a hundred atoms; the dissociation of the largest complex is shown in Figure~\ref{WCCR10-time}. Accurately calculating the ZPE, enthalpy and entropy corrections from approximate Hessians of these dissociation energies is challenging, since not only the number of bonds but also the number of molecules are subject to change during the reactions, accompanied by extensive scrambling of low-frequency modes which contribute substantially to the reaction entropies.
	
	Before we discuss the accuracy of the results, we first present results on the time and memory consumption of O1NumHess. As evident in Figure~\ref{WCCR10-time}(b), O1NumHess always requires fewer gradients than the number of gradients required by the conventional double-sided algorithm ($6N_{\mathrm{atom}}$); in fact, by construction, O1NumHess requires at most $3N_{\mathrm{atom}}-4$ gradients, as any larger number of gradients would result in an over-complete set of displacement directions. One would thus predict that O1NumHess exhibits a time-wise speedup factor of at least 2 compared to the conventional algorithm, which is indeed observed for all but the smallest systems (Figure~\ref{WCCR10-time}(c)). More importantly, however, the number of gradients required by O1NumHess grows more and more slowly as a function of system size, and reaches a plateau of about 120 for sufficiently large systems (Figure~\ref{WCCR10-time}(b)). This is in line with our prediction that O1NumHess only needs $O(1)$ gradients, and results in a linear speedup compared to the conventional algorithm (Figure~\ref{WCCR10-time}(c)).
	
	Interestingly, for most molecules in WCCR10, O1NumHess is even faster than the corresponding analytic Hessian calculation, and the acceleration ratio increases as a function of system size (Figure~\ref{WCCR10-time}(d)), suggesting that O1NumHess may be useful even when analytic Hessians are available. This can be partly attributed to the limited memory budget for the analytic Hessian calculations, which necessitates solving the CP-SCF equations in multiple batches (up to 12). The good parallelization efficiency of O1NumHess may be also playing a role here. We would also like to stress that there may be room for further efficiency improvement in BDF's analytic Hessian, and the high efficiency of O1NumHess is accompanied by somewhat larger numerical errors (see next paragraph). Nevertheless, O1NumHess consumes one order of magnitude less memory than analytic Hessian calculations (Figure~\ref{WCCR10-time}(e)), suggesting that even in the presence of more efficient analytic Hessian implementations, O1NumHess may still be advantageous over analytic Hessian when memory is limited.
	
\begin{figure}[!]
		\centering
		\includegraphics[width=0.8\textwidth]{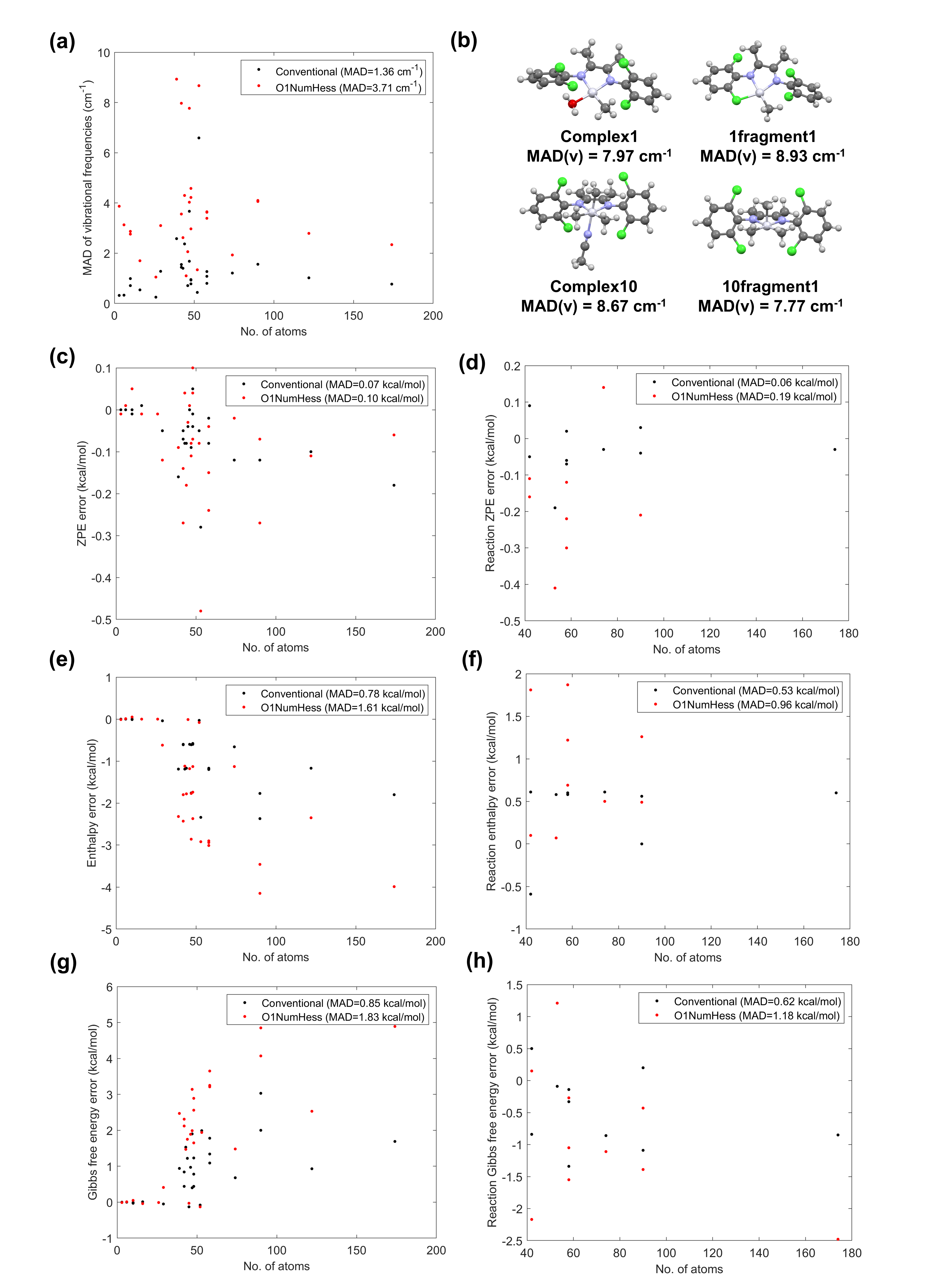}
		\caption{Accuracy of O1NumHess on the WCCR10 set.  (a) MADs of the vibrational frequencies of each molecule, compared to analytic Hessians; (b) the four molecules that have the largest frequency errors; errors of absolute (c) and reaction (d) ZPEs, absolute (e) and reaction (f) enthalpies, and absolute (g) and reaction (h) Gibbs free energies compared to analytic Hessians.}
		\label{WCCR10-accuracy}
	\end{figure}

	With the encouraging results in computational timings, we now investigate the accuracy of O1NumHess on the WCCR10 set. Of the 30 molecules in the WCCR10 set, 26 have vibrational frequency errors below 5 cm$^{-1}$ (Figure~\ref{WCCR10-accuracy}(a)); the remaining four outliers all contain platinum (Figure~\ref{WCCR10-accuracy}(b)), suggesting that the errors may be mostly due to the large DFT grid error associated with the heavy platinum atom, which is magnified by the O1NumHess calculation procedure. Overall, O1NumHess gives somewhat larger errors than the conventional numerical Hessian algorithm, but which are still one order of magnitude smaller than the frequency errors of typical density functionals. O1NumHess systematically underestimates the ZPE and enthalpy corrections (Figure~\ref{WCCR10-accuracy}(c,e)), and overestimates the Gibbs free energy corrections (Figure~\ref{WCCR10-accuracy}(g)); however the same trend is seen in the results of the conventional seminumerical Hessian algorithm, and the errors of O1NumHess are only around twice the errors of the conventional algorithm. Thanks to systematic error cancellation, the reaction enthalpies and Gibbs free energies (but not the ZPEs) have even smaller errors than the absolute enthalpies and Gibbs free energies of the individual reactants and products (Figure~\ref{WCCR10-accuracy}(d,f,h)), and remain around twice of the conventional algorithm's errors. The reaction Gibbs free energy MAD, 1.18 kcal/mol, is well within the errors of typical density functionals for this benchmark set (5-10 kcal/mol\cite{WCCR10}), smaller than the difference of the QRRHO and rigid rotor harmonic oscillator (RRHO) reaction Gibbs free energies for some systems (up to 4.3 kcal/mol\cite{QRRHO}), and comparable to the change of Gibbs free energy with respect to molecular rotation due to integration grid error (1-2 kcal/mol even with relatively large integration grids\cite{Wheeler2019grid}). We therefore consider it safe to use O1NumHess to study the ZPE-corrected reaction energies, reaction enthalpies and reaction Gibbs free energies, at least for reactions that are similar to those studied herein.
	
	\subsection{Large non-covalent complexes: the S30L-CI set}
	
\begin{figure}[!]
		\centering
		\includegraphics[width=\textwidth]{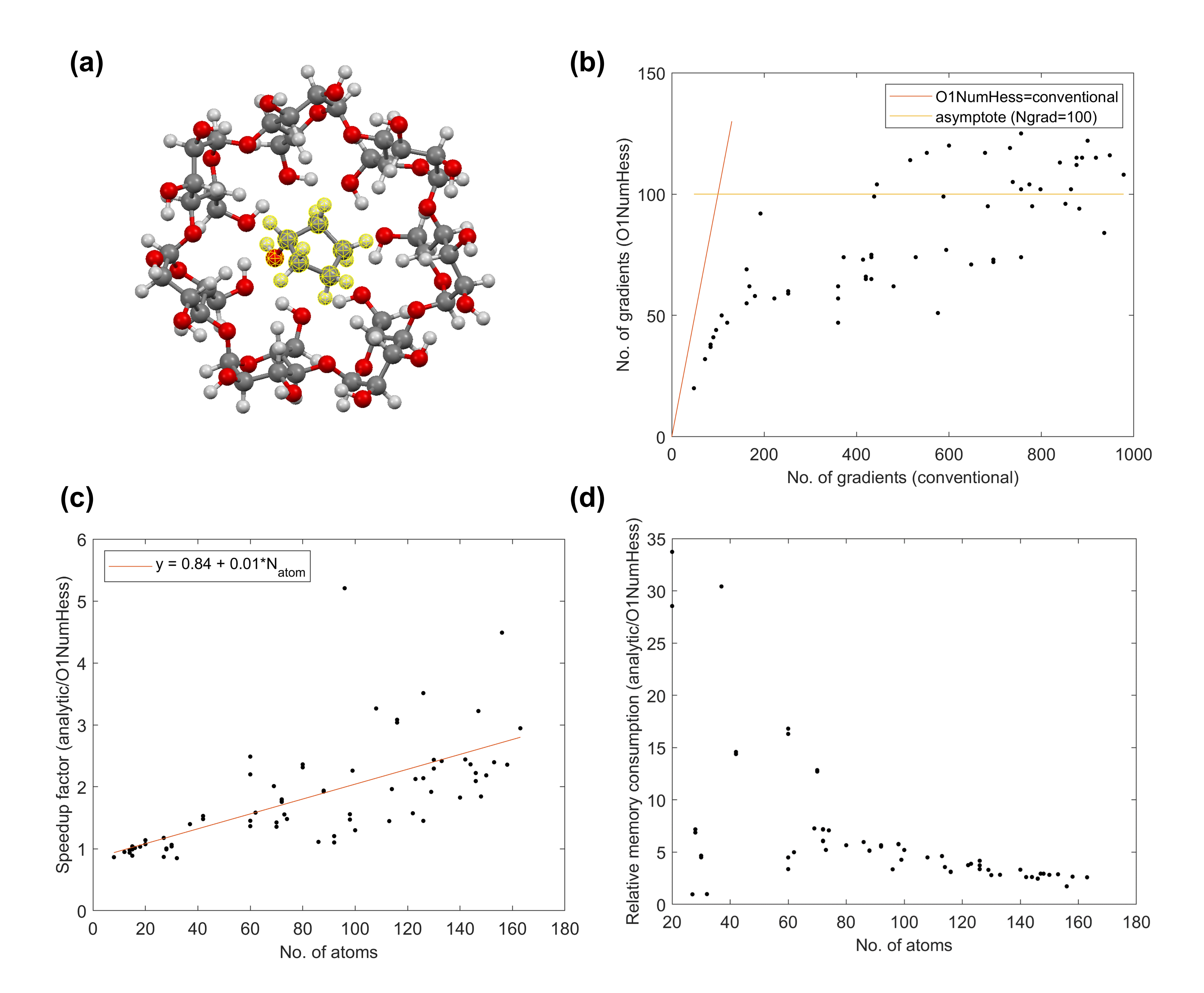}
		\caption{Efficiency of O1NumHess on the S30L-CI set. (a) A representative complex in the S30L-CI set; (b) numbers of gradients required by O1NumHess, compared to those required by the conventional double-sided seminumerical Hessian algorithm; (c) speedup factors of O1NumHess compared to analytic Hessian; (d) relative memory usage of analytic Hessian compared to O1NumHess.}
		\label{S30L-CI-time}
	\end{figure}

	Compared to coordination complexes, non-covalent complexes are more challenging for approximate (semi)numerical Hessian algorithms, because they are more sensitive to low frequency modes. We chose S30L-CI\cite{S30L} as a set of prototypical large non-covalent complexes (the largest one, ``AB19'', is shown in Figure~\ref{S30L-CI-time}(a)), upon which we benchmarked the accuracy of O1NumHess for the binding Gibbs free energies. Note that many of the systems contain one or more chloride or sodium ions to neutralize their charges, which further increases the number of low frequency modes. Due to the high computational cost, we do not report timing results of the conventional Hessian calculations for this set.
	
	Similar to the WCCR10 set, the numbers of gradients required by O1NumHess for the largest S30L-CI systems saturate at about 100 on average, with a maximum of 124 gradients for the complex ``AB3'' (Figure~\ref{S30L-CI-time}(b)). O1NumHess continues to provide increasing speedup factors compared to analytic Hessians, with a slope and intercept similar to those of WCCR10 (Figure~\ref{S30L-CI-time}(c)). Although O1NumHess still provides a memory saving for almost all S30L-CI systems compared to analytic Hessians, the improvement ratio is smaller than that of WCCR10, particularly for large systems where the memory saving stabilizes at two to three fold (Figure~\ref{S30L-CI-time}(d)); this may be due to the higher memory consumption of the IEFPCM gradient compared to the gas phase gradient. Except for the somewhat larger ZPE errors, O1NumHess gives similar frequency and thermochemical property errors on the S30L-CI set compared to the WCCR10 set (Figure~\ref{S30L-CI-time}(a,c-h)), which is gratifying given that the S30L-CI set focuses on non-covalent complexation energies and thus should be more sensitive to errors of low-frequency modes. In particular, the MAD of complexation Gibbs free energies (0.92 kcal/mol) is much smaller than the complexation energy MAD of the most accurate level of theory tested in the S30L-CI paper ($\omega$B97X-D3/def2-QZVP'; 2.1 kcal/mol\cite{S30L}), suggesting that using O1NumHess should not significantly deteriorate non-covalent complexation Gibbs free energies results in typical DFT calculations. Due to the absence of heavy elements in the S30L-CI set, the frequency error of the S30L-CI set is rather uniform, and the largest errors occur in molecules with very large conjugated systems (Figure~\ref{S30L-CI-time}(b)), paralleling our previous observations in the \ce{C32H34} model system. The complexation of host molecule ``A7'' with its guest molecule (``B7'', which is similar to ``A7'' but has only five phenylene groups instead of eight) is also the complexation reaction that has the largest Gibbs free energy error (-2.54 kcal/mol), but which is still much smaller in magnitude than the complexation Gibbs free energy itself (-25.1 kcal/mol). In sum, we conclude that O1NumHess is equally reliable for non-covalent complexation energies compared to reaction energies, and can be safely used to study processes where the entropy contributions of low frequency modes are important.
	
	\begin{figure}[!]
		\centering
		\includegraphics[width=0.8\textwidth]{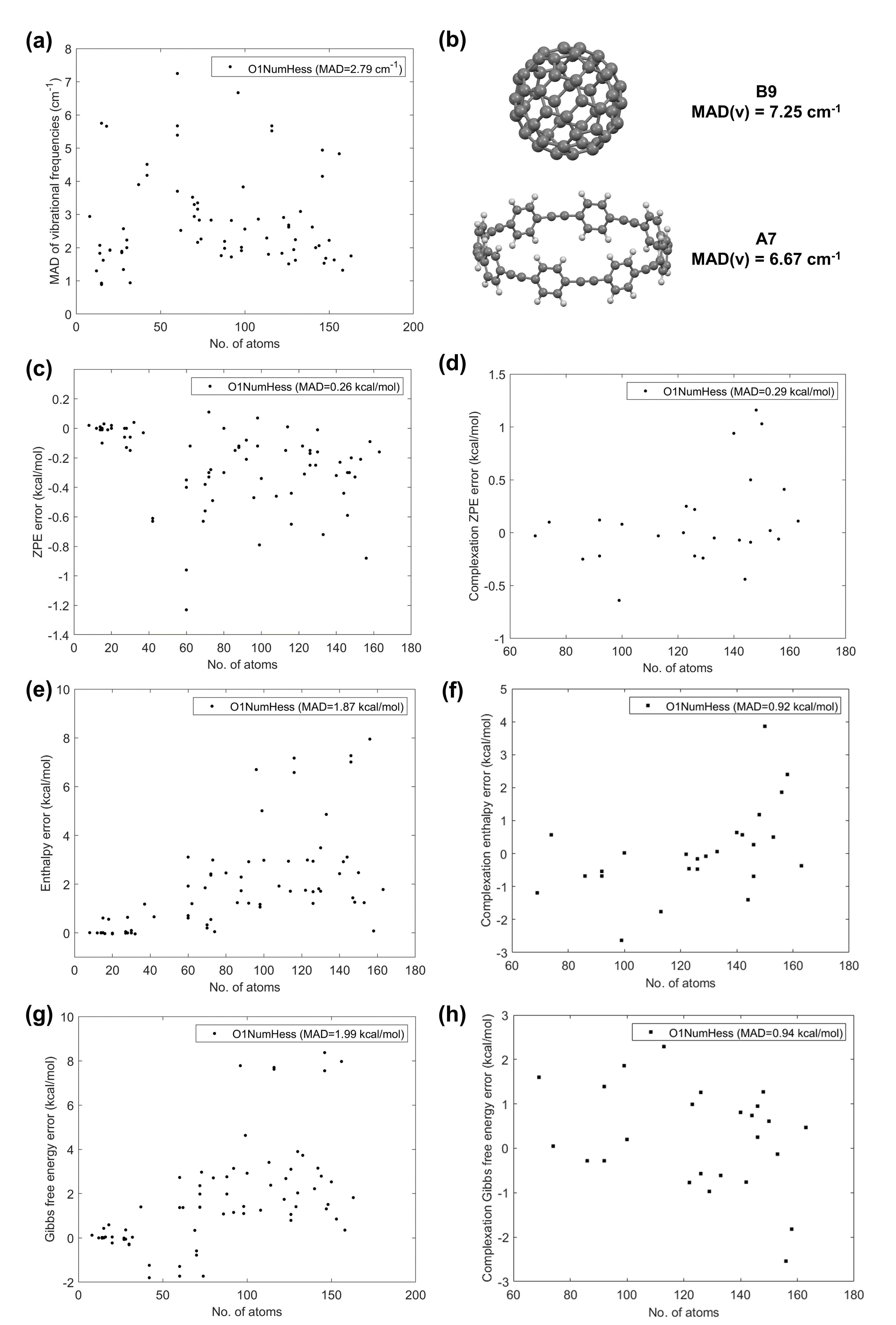}
		\caption{Accuracy of O1NumHess on the S30L-CI set.  (a) MADs of the vibrational frequencies of each molecule, compared to analytic Hessians; (b) the two molecules that have the largest frequency errors; errors of absolute (c) and reaction (d) ZPEs, absolute (e) and reaction (f) enthalpies, and absolute (g) and reaction (h) Gibbs free energies compared to analytic Hessians.}
		\label{S30L-CI-accuracy}
	\end{figure}
	
	\section{Conclusions and outlook}
	
	We have developed a general approach, O1NumHess, for estimating the Hessian of a molecular system using gradients at an asymptotically constant ($O(1)$) number of displaced geometries, taking advantage of the ODLR property of Hessians. This represents a logarithmic speedup compared to the best known algorithm\cite{Aspuru2015} for local Hessians, and a linear speedup for Hessians that lack locality (in which case no known algorithms exist that require fewer than $O(N_{\mathrm{atom}})$ gradients for reproducing the nonlocal part of the Hessian). Benchmark results show that O1NumHess gives numerical errors that are merely twice those given by the conventional double-sided seminumerical differentiation algorithm, at a cost of only about 100-120 gradients for large systems, compared to $6N_{\mathrm{atom}}$ gradients needed by the conventional double-sided algorithm. For all systems studied herein, O1NumHess is faster than the conventional seminumerical Hessian algorithm, and a linear increase in the speedup factor with respect to system size is observed. Therefore, we recommend using O1NumHess in any case where the analytic Hessian is not available, or not affordable due to memory restrictions. In particular, O1NumHess is expected to be an important tool for the calculation of excited state Hessians, since analytic TDDFT Hessians are not available in many popular programs. Finally, we have shown that O1NumHess may even be faster than some analytic Hessian implementations, suggesting that O1NumHess may be useful even when an analytic Hessian calculation is affordable. However, it must be kept in mind that the speedup of O1NumHess is accompanied by a larger numerical error.
	
	Although currently the \verb|O1NumHess_QC| library only has interfaces with BDF and ORCA, we strongly encourage users to implement its interfaces with other programs, or integrate \verb|O1NumHess| and \verb|O1NumHess_QC| into various electronic structure programs. Perhaps a more interesting possibility is the application of \verb|O1NumHess| to the calculation of Hessians other than the nuclear Hessian, such as the orbital Hessian, or even Hessians that are unrelated to computational chemistry. These are facilitated by the fact that the \verb|O1NumHess| code accepts general gradient functions and does not assume that the gradients are the nuclear derivatives of the electronic energy.
	
	Finally, Hessians are not the only matrices that possess the ODLR property. It is long known in the mathematics community\cite{Boulle2024} that the ODLR phenomenon can be observed in the matrix discretizations of sufficiently smooth and fast-decaying operators (the so-called Calderon-Zygmund operators\cite{CalderonZygmund}). In particular, this includes the matrix representations of the Coulomb operator, which is the basis of the renowned fast multipole method (FMM)\cite{FMM1,FMM2,FMM3,FMM4,FMM5,FMM6}. It is hoped that the present work will inspire more algorithms that exploit the ODLR property of other matrices, for example the density matrix\cite{ERG}, the Fock matrix and the electron repulsion integral tensor\cite{Xing2020,Xing2020b,H2Pack,BUPO}. Work is ongoing along these directions.
	
	\section*{Acknowledgement}
	
	Z.W. acknowledges the support of the Qilu Young Scholar Program of Shandong University, and would like to acknowledge Prof.~Bingbing Suo (Northwest University) for inspiring discussions.
	
	\appendix
	
	\section{Proper treatment of molecular rotations} \label{appendix:transrot}
	
	Assuming the molecule is not subject to an external electric field or an inhomogeneous medium, its translational frequencies are zero, so that we can include the three translational modes in the list of displacement directions and set the respective gradients as zero. However, this is not true for rotations: the frequencies along the rotational modes are generally only zero when the molecule is at its equilibrium geometry. Specifically, when a molecule has a non-vanishing gradient $\mathbf{g}_0$, and is rotated by an infinitesimal angle $d\theta$ around an axis $\mathbf{n}=\{n_x,n_y,n_z\}, |\mathbf{n}|=1$, the gradient changes by
	\begin{equation}
		d\mathbf{g}_0 = (\mathbf{n}\times\mathbf{g}_0)d\theta.
	\end{equation}
	Although the gradient merely changes its direction but does not change its magnitude, it still contributes to the Hessian because the Hessian is the derivative of the gradient vector (not its amplitude) with respect to the nuclear coordinates. Meanwhile, such a rotation changes the nuclear coordinates by
	\begin{equation}
		d\xi_i = (\mathbf{n}\times(\xi_i-\xi_0))d\theta,
	\end{equation}
	where $\xi_0$ is the barycenter of the molecule, but calculated assuming that all atoms in the molecule have the same mass. The latter ensures that O1NumHess gives the same Hessian for different isotopologues of the same molecule, which is convenient for studies of isotope substitution effects. The scaled gradient $g_{ij}$ (where $i$ is a rotation mode around the axis $\mathbf{n}$), which is the ratio of the gradient change with the norm of the displacement, can thus be calculated as
	\begin{equation}
		g_{ij} = \frac{d\mathbf{g}_0}{\sqrt{\sum_j(d\xi_{j})^2}} = \frac{\mathbf{n}\times\mathbf{g}_0}{\|\mathbf{n}\times(\xi_i-\xi_0)\|}. \label{grot}
	\end{equation}
	Eq.~\eqref{grot} allows us to save 3 gradient evaluations (2 for linear molecules) for molecules that are not at their equilibrium geometries, even though their Hessians have non-zero eigenvalues along the rotational modes.
	
	The three rotational axes $\mathbf{n}$ (two for linear molecules) are chosen as the eigenvectors of the moment of inertia tensor of the molecule (as is usually done), but where the masses of all atoms are treated as equal.
	
	\section{The model Hessian} \label{appendix:Swart}
	
	The model Hessian proposed by Swart et al.\cite{Swart}~is a cheap Hessian computed from empirical force constants of bond, angle and dihedral degrees of freedom. It is used as the default initial Hessian for geometry optimizations in ORCA. As it is even cheaper than molecular mechanics Hessians (due to the absence of Coulomb and van der Waals interactions), we have chosen Swart et al.'s model Hessian as an initial approximation of the real Hessian during generation of the displacement directions, after modifying the expressions to make them more suitable for complex bonding patterns.
	
	Our approach starts from calculating a diagonal Hessian under the redundant internal coordinate basis, followed by transforming the Hessian to the Cartesian coordinates. All pairs of atoms are defined as bonds; this differs from the original Swart et al.'s Hessian, where only atom pairs where $\rho_{AB}\ge0.3$ are treated as bonds.
	$\rho_{AB}$ is defined as ($r_A^{\mathrm{cov}}$ is the covalent radius of atom $A$\cite{Pyykko_covrad})
	\begin{equation}
		\rho_{AB} = \exp(1-r_{AB}/(r_A^{\mathrm{cov}}+r_B^{\mathrm{cov}})).
	\end{equation}
	Such a modification removes the need of adding dihedral and improper dihedral angles to the list of redundant coordinates. The force constants of bonds are calculated in the same way as in Swart et al.'s implementation, as
	\begin{equation}
		k_{AB} = 0.35\rho_{AB}^3.
	\end{equation}
	
	Atom triplets $ABC$ where $\rho_{AB}\rho_{BC}\ge0.09$ are defined as angles. When $|\cos \angle ABC| \le 0.8$, the angle $\angle ABC$ is called a normal angle, and its force constant is
	\begin{equation}
		k_{ABC} = 0.075(\rho_{AB}\rho_{BC}(0.12+0.88\sin\angle ABC))^2,
	\end{equation}
	which differs from Swart et al.'s original implementation by a factor of 0.5. Angles for which $\cos \angle ABC > 0.8$ are called close-to-zero angles, whose force constants are scaled down by a factor $(1-\sigma_{ABC})^2$, where
	\begin{equation}
		\sigma_{ABC} = \left(1 - \left(\frac{1-|\cos \angle ABC|}{0.2}\right)^2\right)^2.
	\end{equation}
	Thus, when the angle approaches $0^{\circ}$, the force constant of ``linear angle 1'' smoothly approaches zero.
	
	Angles for which $\cos \angle ABC < -0.8$ are linear angles. It is well-known from the common experiences of geometry optimization that each linear angle requires two internal coordinates for a numerically stable description, ``linear angle 1'' and ``linear angle 2''. When the angle $ABC$ is close to but not exactly $180^{\circ}$, ``linear angle 1'' is equal to the normal angle bending mode, while ``linear angle 2'' is a rigid rotation of the $ABC$ moiety around an axis that is parallel to the line connecting $A$ and $C$. In the limit $\angle ABC\to 180^{\circ}$, ``linear angle 1'' and ``linear angle 2'' reduce to the two independent angle bending motions of the linear angle $ABC$. The force constants for ``linear angle 1'' and ``linear angle 2'' are given respectively by
	\begin{equation}
		k_{ABC}^{\mathrm{lin1}} = k_{ABC},
	\end{equation}
	\begin{equation}
		k_{ABC}^{\mathrm{lin2}} = \sigma_{ABC}^2k_{ABC}.
	\end{equation}
	When the angle approaches $180^{\circ}$, the force constant of the ``linear angle 1'' remains non-zero, while the force constant of ``linear angle 2'' smoothly increases from zero and reaches the same value as that of ``linear angle 1'' in the limit of an exactly linear angle.
	The above treatments have been verified to work well in various molecules that contain linear, near-linear or close-to-zero angles.
	
	One interesting possibility is to use our modified model Hessian as the initial Hessian of geometry optimizations. It remains to be seen whether this would provide better convergence behavior compared to the original model Hessian by Swart et al.
	
	\clearpage
	\newpage
	
	\bibliographystyle{apsrev4-1}
	\bibliography{O1NumHess}
	
\end{document}